\documentclass[pdflatex,sn-mathphys-num]{sn-jnl}% Math and Physical Sciences Numbered Reference Style 
\usepackage{graphicx}%
\usepackage{multirow}%
\usepackage{amsmath,amssymb,amsfonts}%
\usepackage{amsthm}%
\usepackage{mathrsfs}%
\usepackage[title]{appendix}%
\usepackage{xcolor}%
\usepackage{textcomp}%
\usepackage{manyfoot}%
\usepackage{booktabs}%
\usepackage{algorithm}%
\usepackage{algorithmicx}%
\usepackage{algpseudocode}%
\usepackage{listings}%
\usepackage{float}
\floatstyle{plaintop}
\restylefloat{table}
\usepackage{tabularx}
\usepackage{adjustbox}
\usepackage{anyfontsize}
\usepackage[acronym]{glossaries}
\setkeys{glslink}{hyper=false}
\newacronym{mrclean}{MR CLEAN}{Multicenter Randomized Clinical trial of Endovascular treatment for Acute ischemic stroke in the Netherlands}
\newacronym{isles}{ISLES}{Ischaemic Stroke Lesion Segmentation}
\newacronym{ct}{CT}{Computed Tomography}
\newacronym{ncct}{NCCT}{Non-contrast Computed Tomography}
\newacronym{ctp}{CTP}{CT Perfusion}
\newacronym{cta}{CTA}{CT Angiography}
\newacronym{mri}{MRI}{Magnetic Resonance Imaging}
\newacronym{mpmri}{mpMRI}{Multiparametric MRI}
\newacronym{mra}{MRA}{Magnetic Resonance Angiography}
\newacronym{mrp}{MRP}{Magnetic Resonance Perfusion}
\newacronym{pet}{PET}{Positron Emission Tomography}
\newacronym{dwi}{DWI}{Diffusion-weighted MRI}
\newacronym{cbf}{CBF}{Cerebral Blood Flow}
\newacronym{cbv}{CBV}{Cerebral Blood Volume}
\newacronym{ttp}{TTP}{Time To Peak}
\newacronym{mtt}{MTT}{Mean Transit Time}
\newacronym{flair}{FLAIR}{Fluid-attenuated Inversion Recover}
\newacronym{adc}{ADC}{Apparent Diffusion Coefficient}
\newacronym{pwi}{PWI}{Perfusion-weighted Imaging}
\newacronym{dsa}{DSA}{Digital Subtraction Angiography}
\newacronym{tmax}{Tmax}{Time-to-Maximum}
\newacronym{dscpwi}{DSC-PWI}{Dynamic Susceptibility
Contrast PWI} 

\newacronym{hu}{HU}{Hounsfield Unit}
\newacronym{evt}{EVT}{Endovascular Therapy}
\newacronym{ais}{AIS}{Acute Ischaemic Stroke}
\newacronym{is}{IS}{Ischaemic Stroke}
\newacronym{hs}{HS}{Haemorrhagic Stroke}
\newacronym{ht}{HT}{Haemorrhagic Transformation }
\newacronym{cad}{CAD}{Computer Aided Diagnosis}
\newacronym{mrs}{mRS}{modified Rankin Scale}
\newacronym{lvo}{LVO}{Large Vessel Occlusion}
\newacronym{mca}{MCA}{Middle Cerebral Artery}
\newacronym{aspects}{ASPECTS}{Alberta Stroke Programme Early CT Score}
\newacronym{nihss}{NIHSS}{National Institutes of Health Stroke
Scale}
\newacronym{bi}{BI}{Barthel Index}
\newacronym{sich}{sICH}{Symptomatic Intracerebral Haemorrhage}
\newacronym{ich}{ICH}{Intracerebral Haemorrhage}
\newacronym{mtici}{mTICI}{modified Thrombolysis in Cerebral Infarction}
\newacronym{tici}{TICI}{Thrombolysis in Cerebral Infarction}

\newacronym{bl}{BL}{Baseline}
\newacronym{fu1w}{FU1W}{Follow-up 1-week}
\newacronym{fu24h}{FU24H}{Follow-up 24-hour}
\newacronym{2d}{2D}{2-dimensional}
\newacronym{3d}{3D}{3-dimensional}

\newacronym{dl}{DL}{Deep Learning}
\newacronym{ml}{ML}{Machine Learning}
\newacronym{cnn}{CNN}{Convolutional Neural Network}
\newacronym{cae}{CAE}{Convolutional Auto Encoder}
\newacronym{nlp}{NLP}{Natural Language Processing}
\newacronym{dnn}{DNN}{Deep Neural Network}
\newacronym{ann}{ANN}{Artificial Neural Network}
\newacronym{cnns}{CNNs}{Convolutional Neural Networks}
\newacronym{lstm}{LSTM}{Long Short-Term Memory}
\newacronym{rbm}{RBM}{Restricted Boltzmann Machines}
\newacronym{gcn}{GCN}{Graph Convolutional Network}
\newacronym{rfnn}{RFNN}{structured Receptive Field Network}
\newacronym{gradcam}{Grad-CAM}{Gradient-weighted Class Activation Map}
\newacronym{cc}{CC}{Correlation Coefficient}
\newacronym{cgan}{CGAN}{Conditional Generative Adversarial Network}

\newacronym{se}{SE}{Squeeze and Excitation}
\newacronym{cse}{cSE}{Channel \acrshort{se}}
\newacronym{sse}{sSE}{Spatial \acrshort{se}}
\newacronym{fcn}{FCN}{Fully Connected Network}
\newacronym{fc}{FC}{Fully Connected}
\newacronym{bn}{BN}{Batch Normalisation}
\newacronym{in}{IN}{Instance Normalisation}
\newacronym{ln}{LN}{Layer Normalisation}
\newacronym{relu}{ReLU}{Rectified Linear Unit}
\newacronym{lrelu}{LeakyReLU}{Leaky Rectified Linear Unit}
\newacronym{gap}{GAP}{Global Average Pooling}

\newacronym{fn}{FN}{False Negative}
\newacronym{fp}{FP}{False Positive}
\newacronym{tn}{TN}{True Negative}
\newacronym{tp}{TP}{True Positive}

\newacronym{lr}{LR}{Logistic Regression}
\newacronym{lir}{LR}{Linear Regression}
\newacronym{rf}{RF}{Random Forest}
\newacronym{dt}{DT}{Decision Tree}
\newacronym{bt}{BT}{Boosted Trees}
\newacronym{knn}{k-NN}{k-Nearest Neighbors}
\newacronym{svm}{SVM}{Support Vector Machine}
\newacronym{rnn}{RNN}{Recurrent Neural Network}
\newacronym{svr}{SVR}{Support Vector Regression}
\newacronym{mlp}{MLP}{Multi Layer Perceptron}
\newacronym{xgb}{XGB}{eXtreme Gradient Boosting}
\newacronym{nb}{NB}{Naive Bayes}
\newacronym{pca}{PCA}{Principal Component Analysis}

\newacronym{ae}{AE}{Auto-encoder}
\newacronym{vae}{VAE}{Variational Autoencoder}
\newacronym{gan}{GAN}{Generative Adversarial Network}
\newacronym{wgan}{WGAN}{Wasserstein GAN}
\newacronym{vit}{ViT}{Vision Transformer}

\newacronym{mse}{MSE}{Mean Square Error}
\newacronym{mae}{MAE}{Mean Absolute Error}
\newacronym{ssim}{SSIM}{Structural Similarity}
\newacronym{mssim}{MS-SSIM}{Multi-scale Structural Similarity}
\newacronym{fl}{FL}{Focal Loss}
\newacronym{pl}{PL}{Perceptual Loss}
\newacronym{ce}{CE}{Cross Entropy}
\newacronym{bce}{BCE}{Binary Cross Entropy}
\newacronym{dicel}{DL}{Dice Loss}
\newacronym{dice}{DSC}{Dice Similarity Coefficient}
\newacronym{auc}{AUC}{Area Under the Curve}
\newacronym{acc}{ACC}{Accuracy}
\newacronym{roc}{ROC}{Receiver Operating Characteristic}
\newacronym{tpr}{TPR}{True Positive Rate}
\newacronym{fpr}{FPR}{False Positive Rate}
\newacronym{icc}{ICC}{Intraclass Correlation Coefficient}

\newacronym{ci}{CI}{Confidence Interval}

\newacronym{mhsa}{MHSA}{Multi-Head Self-Attention}
\newacronym{sa}{SA}{Self-Attention}

\newacronym{cv}{CV}{Computer Vision}

\newacronym{end}{END}{Early Neurological Deterioration}
\newacronym{rtpa}{rtPA}{recombinant tissue-type Plasminogen Activator}
\newacronym{tpa}{tPA}{tissue Plasminogen Activator}
\newacronym{psci}{PSCI}{Post-Stroke Cognitive Impairment}

\newacronym{loocv}{LOOCV}{Leave-One-Out Cross-Validation}
\newacronym{fcv}{cv}{Cross-Validation}

\newacronym{nr}{N/R}{Not Reported}
\newacronym{ns}{N/S}{Not Specified}

\newacronym{sop}{SOP}{Stroke Outcome Prediction}

\usepackage[hyperpageref]{backref}

\usepackage{pifont}

\newcommand{\cmark}{\textcolor{black!90}{\ding{51}}}
\newcommand{\xmark}{\textcolor{black!50}{\ding{55}}}

\usepackage[normalem]{ulem}
\definecolor{Gray}{gray}{0.9}

\DeclareRobustCommand{\eg}{\emph{e.g. }}

\DeclareRobustCommand{\ie}{\emph{i.e. }}

\DeclareRobustCommand{\etal}{\emph{et al. }}

\newcommand\tab[1][1cm]{\hspace*{#1}}
\begin{document}

\title[Article Title]{Automatic Prediction of Stroke Treatment Outcomes: Latest Advances and Perspectives}


\author*[1]{\fnm{Zeynel A.} \sur{Samak}}\email{zsamak@adiyaman.edu.tr}

\author[3,4]{\fnm{Philip} \sur{Clatworthy}}\email{phil.clatworthy@bristol.ac.uk}

\author[2]{\fnm{Majid} \sur{Mirmehdi}}\email{m.mirmehdi@bristol.ac.uk}

\affil*[1]{\orgdiv{Department of Computer Engineering}, \orgname{Adiyaman University}, \orgaddress{\city{Adiyaman}, \postcode{02040}, \country{Türkiye}}}

\affil[2]{\orgdiv{School of Computer Science}, \orgname{University of Bristol}, \orgaddress{\city{Bristol}, \postcode{BS8 1UB},  \country{UK}}}

\affil[3]{\orgdiv{Translational Health Sciences}, \orgname{University of Bristol}, \orgaddress{ \city{Bristol}, \postcode{BS8 1UD},  \country{UK}}}

\affil[4]{\orgdiv{Stroke Neurology, Southmead Hospital}, \orgname{North Bristol NHS Trust}, \orgaddress{\street{Street}, \city{Bristol}, \postcode{BS8 1UD},  \country{UK}}}

% \affil[3]{\orgdiv{Department}, \orgname{Organization}, \orgaddress{\street{Street}, \city{City}, \postcode{610101}, \state{State}, \country{Country}}}


\abstract{Stroke is a major global health problem that causes mortality and morbidity. Predicting the outcomes of stroke intervention can facilitate clinical decision-making and improve patient care. Engaging and developing deep learning techniques can help to analyse large and diverse medical data, including brain scans, medical reports and other sensor information, such as EEG, ECG, EMG and so on.  Despite the common data standardisation challenge within medical image analysis domain, the future of deep learning in stroke outcome prediction lie in using multimodal information, including final infarct data, to achieve better prediction of long-term functional outcomes. This article provides a broad review of recent advances and applications of deep learning in the prediction of stroke outcomes, including (i) the data and models used, (ii) the prediction tasks and measures of success, (iii) the current challenges and limitations, and (iv) future directions and potential benefits.  This comprehensive review aims to provide researchers, clinicians, and policy makers with an up-to-date understanding of this rapidly evolving and promising field. }




\maketitle

\section{Introduction}\label{sec1}

\label{sec:intro}

A stroke is a critical medical event that occurs when the blood supply to the brain is interrupted. This disrupts oxygen delivery, leading to tissue damage and neurological impairment. The lack of blood and oxygen damages brain cells, ultimately causing neuronal death. This can lead to permanent brain damage, long-term functional disability or even death. Several key factors increase the risk of stroke, including high blood pressure, diabetes, high cholesterol, smoking, heart disease, obesity and a family history of stroke. Age, gender and ethnicity can also contribute \cite{boehme2017stroke}. 

Stroke is the second leading cause of death and the third leading cause of disability worldwide, affecting approximately 15 million people annually \cite{zhao2019lifetime}, according to the World Health Organization \cite{WHO2020}.  The well-known phrase "time is brain" \cite{saver2006time} highlights the critical importance of acting quickly in the assessment and treatment of stroke. Delays in treatment can lead to brain cell death from lack of blood flow and oxygen, and research suggests that approximately two million neurons die every minute an ischaemic stroke goes untreated \cite{saver2006time}. A comprehensive analysis of registry data revealed that each minute delay in intravenous thrombolysis administration increases the risk of early post-treatment intracerebral hemorrhage, disability, and mortality \cite{darehed2020hospital}. Clinicians therefore aim to rapidly use all available data sources, including imaging and clinical information, to determine the most appropriate treatment option as quickly as possible. 

The outcome for stroke patients is influenced by a number of factors, including the type, location and size of the stroke, the time elapsed before treatment and the rehabilitation interventions received \cite{sacco2013updated}. Despite significant advances in understanding the mechanisms of stroke and developing effective treatments, {like thrombectomy (mechanically removing blood clots, \gls{evt}) and thrombolysis (dissolving blood clots with medication)}, predicting patient outcomes and selecting the most appropriate treatment remains challenging because it involves complex interactions between multiple variables that may change over time \cite{heo2019machine}. 
Consequently, developing an automated tool that uses available imaging and patient information to predict treatment outcomes would improve clinical decision-making and improve the efficacy of ischaemic stroke treatment.

%\mmn{In the interest of word count, I suggest we skip these bullet points for now\\}\zsn{Ok}
%\begin{itemize}
%    \item \zs{Prognostic information allows clinicians to select the most appropriate treatment strategies for individual treatment plans. If a particular therapy suggests a low chance of success, clinicians can explore alternative options or refine the existing plan to better benefit the patient.}
%    \item \zs{By tailoring treatment plans to each patient's specific needs and predicted outcome, clinicians can potentially improve overall patient outcomes and minimise the risk of complications.}
%   \item \zs{Identifying the patients most likely to benefit from a treatment enables targeted recruitment to clinical trials. This ensures that trials enrol the most appropriate patient population, potentially leading to more successful outcomes and faster development of effective therapies.}
%   \item \zs{Identification of patients with a higher likelihood of poor outcomes enables clinicians to allocate resources effectively, prioritising those who are more likely to benefit from treatment and improving the overall efficiency of healthcare delivery.} 
%\end{itemize}

The early prediction of stroke outcome and delivery of the appropriate treatment is critical since the ischaemic penumbra, the region of brain tissue surrounding the infarct that can still be saved, has a limited window for successful intervention \cite{Karthik2020}. {To address this, researchers categorise stroke analysis into distinct phases and tasks (see also Figure \ref{fig:time_framel}) to guide treatment decisions \cite{Inamdar2021,Zeng2022}. The initial assessment occurs upon hospital admission to identify brain tissue damage. Subsequent analyses are conducted to evaluate how the tissue is changing in response to treatment and determine longer-term functional outcomes, typically within the first week and at three months.} {Importantly, all predictions within these time-frames must be made after the initial hospital admission scan to help clinicians choose the most appropriate course of treatment.

\begin{figure}[t]
    \centering
    \includegraphics[width=\linewidth]{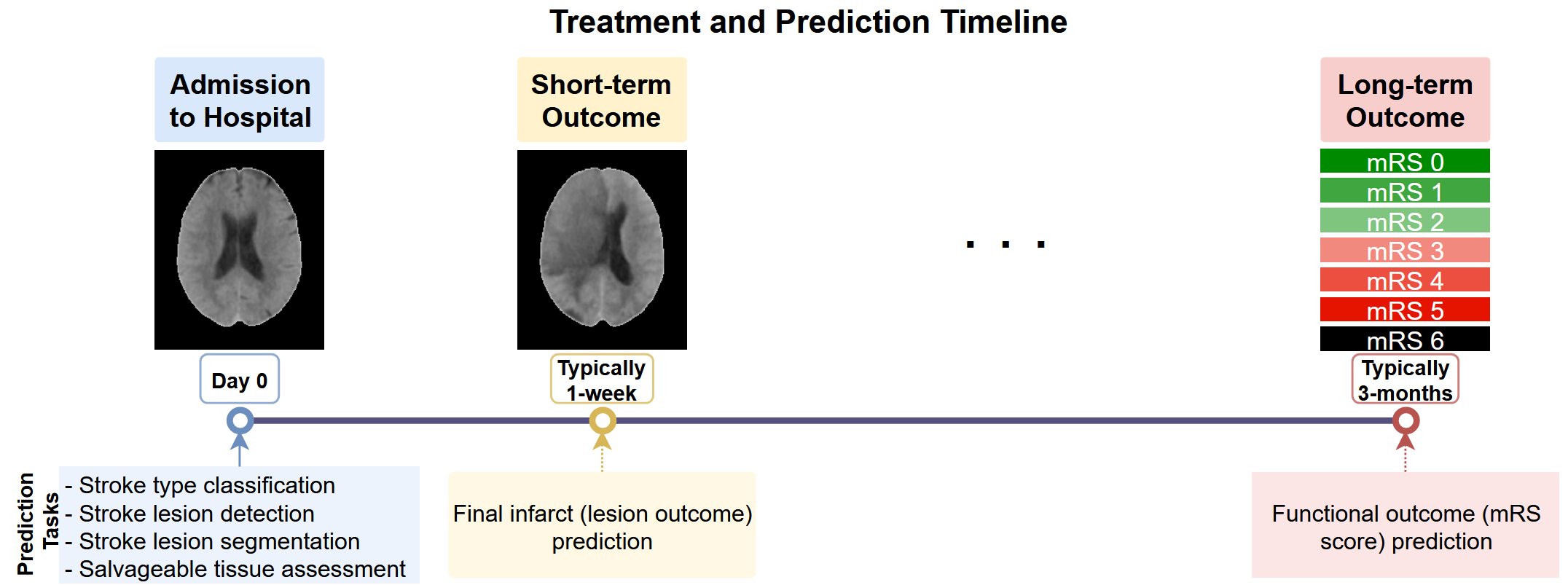}
    \caption{Analysis of stroke time frames and the predicted tasks at each time point. {These predictions are based on initial scans performed upon hospital admission to aid physicians in their clinical decision-making process.}} 
    \label{fig:time_framel}
\end{figure}

To make accurate predictions within these time-frames, researchers use a variety of data sources: brain imaging techniques such as \gls{ct} scans (used for rapid stroke assessment), \gls{mri} (providing detailed anatomical information) and clinical data including patient demographics, blood biomarkers and neurological scores. {Given the requirement of quick reactions to determine diagnosis and perform corrective procedures, clinical and radiological {\it reports} are either not available or feature less in such situations. Hence, there is also a lack of such data for researchers to apply today's enhanced language models \cite{radford2021learning,touvron2023llama} for outcome predictions. However, this is an important area which has scope for development.}}

Early attempts to automate stroke analysis relied on basic techniques such as thresholding and standard statistical methods. These methods included correlation-based analysis \cite{Chawla2009AImages,Rekik2012MedicalAppraisal}, region growing \cite{Rekik2012MedicalAppraisal,gupta2014brain}, and classical \gls{ml} algorithms such as \gls{rf} \cite{mckinley2017fully,bohme2018combining} and \gls{svm} \cite{Maier2014IschemicClassifiers}. In recent years, \gls{ml} and \gls{dl} methods have gained significant popularity due to their effectiveness in various image analysis tasks, such as object detection \cite{lin_focal_2017,cai_cascade_2017}, image segmentation \cite{isensee2017brain,Li2017NotCascade}, and classification \cite{he2016deep,Hu2018Squeeze-and-ExcitationNetworks}. \gls{dl}, in particular via \gls{cnn} \cite{kamnitsas2017ensembles,hilbert2019data, pinto2021combining} and recently Transformers \cite{hatamizadeh2022swin, amador2022predicting,Samak_2023}, has been widely applied to medical image analysis tasks with superior results, including automated ischaemic stroke {lesion classification} \cite{gautam2021towards,neethi2022stroke}, {detection/segmentation} \cite{kamnitsas2017ensembles,pinto2021combining} and {prognosis} \cite{nishi2020deep,Bacchi2019DeepStudy,Samak_2022}. {Most \gls{sop} studies have focused on either predicting the final lesions or predicting functional outcome at 90-days. While  \citet{ernst2017association} demonstrated the 1-week follow-up scan  containing final lesions has a positive impact on predicting the functional outcome, few studies have investigated integrating the final lesion and functional outcome prediction tasks \cite{nishi2019predicting,Samak_2022}.

\begin{table}[t]
\centering
\large
\resizebox{\textwidth}{!}{
    \begin{tabular}{lcclrr}
    \toprule
         \textbf{Study}&  \textbf{Year}&  \textbf{Coverage} &\textbf{Reviewed Topics}& \textbf{\#Papers}&\textbf{\#DL in \acrshort{sop}}\\
         \midrule
         \citet{Sirsat2020}&  2020&   2007-2019& ML in stroke identification, diagnosis, treatment \& prognostication&  39& 1\\
         \midrule
         \citet{Karthik2020}&  2020&  to 2020&Stroke lesion detection and segmentation&  113& 7\\
         \midrule
         \citet{Inamdar2021}&  2021&   2010-2021&Detection and lesion segmentation of neuroimage&  117& 7\\
         \midrule
         % \citet{Yeo2021}&  2021&   N/S&AI applications in stroke&  N/S& 2\\
         % \midrule
         \citet{Mainali2021}&  2021& to 2021 &ML in stroke diagnosis and outcome prediction&  54& 3\\
         \midrule
         \citet{Zeng2022}   & 2022& to early 2022& Functional outcome, reperfusion, and hemorrhagic transformation& 16&4\\
         \midrule
         \citet{Kremers2022}& 2022& to mid-2020& Functional Outcome after EVT and using  \acrshort{mrclean} dataset& 19&2\\
         \midrule
         \citet{Wang2022}& 2022& to mid-2022& ML algorithms for final infarct prediction& 11&7\\
         \midrule
         \citet{Hilbert2022}& 2022& to mid-2022& AI applied in stroke decision support& 65&11\\
         \midrule
         \textbf{Our Review}& \textbf{2024}& \textbf{2015-2024}& \textbf{DL in \acrshort{sop} (final infarct \& functional outcome)}& \textbf{60}&\textbf{60}\\

\bottomrule
    \end{tabular}

}
\caption{Recent {review/survey articles that included} \acrfull{sop}, their coverage, topics reviewed, total number of papers reviewed, including total number of articles related to \acrshort{sop}. }
\label{table:review_list}
\end{table}

Several previous review papers (summarised in Table \ref{table:review_list}) have investigated the application of ML and DL techniques in the domain of brain stroke. 
While these prior reviews have comprehensively addressed stroke classification \cite{Sirsat2020,Mainali2021}, detection \cite{Karthik2020,Inamdar2021}, and segmentation \cite{Karthik2020,Hilbert2022}} and many have focused on classical ML methods \cite{Sirsat2020,Mainali2021,Wang2022}, {the application and progress of \gls{dl} in \gls{sop} has not been coalesced and categorised.}

{In this paper, we} present an overview of the different \gls{dl} approaches that have been  applied to predict the outcome of stroke treatment as the final lesion and functional outcome using only imaging and multimodal data.  We  discuss the strengths and limitations of these techniques and 
critically evaluate the current state of the field, identifying gaps and areas that require further investigation to improve the precision and reliability of \gls{sop}. Additionally, we consider the public datasets that are available for \gls{sop} applications, {as well as list the works for which code {(see Appendix \ref{secA1})} and data are available.} 

The rest of paper is structured as follows. In Section \ref{sec:datasets}, we consider existing, relevant datasets and score measures for stroke outcome prediction. Section \ref{sec:deep_learning_stroke} reviews articles in \gls{sop} using \gls{dl} methods under two categories: final infarct and functional outcome prediction. 
%Section \ref{sec:outcome_mrs} further explores functional outcome prediction, examining methods based on image data only (Section \ref{sec:outcome_image}) and multimodal data (Section \ref{sec:outcome_multimodal}). 
In the Discussion in Section \ref{sec:discussion}, we present our key observations and review current limitations. 
%analyse the implications of the findings and discuss potential future research directions 
Then, in Section \ref{sec:future_directions}, we suggest avenues of further exploration for future directions of research for more efficient stroke outcome prediction. 
%and acknowledge any limitations of this study in Section \ref{sec:limitations}. 
Finally, Section \ref{sec:conclusion} concludes the paper.
%by summarising the key findings and highlighting the potential impact of \gls{dl} for improving stroke treatment outcomes.}

\section{Datasets \& Scores}   \label{sec:datasets}
Publicly available datasets are scarce within the \gls{sop} domain, hence some studies use their own in-house collections. Table \ref{table:datasets} lists only datasets that have been used in at least two studies of \gls{sop} research. While access to most of these datasets requires application and approval, the Ischemic Stroke Lesion Segmentation (ISLES) 2017 \cite{Winzeck2018ISLESMRI} dataset, released for a challenge at the MICCAI \footnote{International Conference on Medical Image Computing and Computer Assisted Intervention - https://miccai.org/} conference, stands out as a publicly available resource, providing valuable data for developing \gls{sop} models and benchmarking for predicting stroke lesion outcome using \gls{mri} data.

\begin{table}[!t]
\centering
\large
\resizebox{\textwidth}{!}{
    \begin{tabular}{lcccrccc}
    \toprule
       \multirow{2}{*}{\textbf{Study}}  & \multirow{2}{*}{\textbf{Year}} & \multirow{2}{*}{\textbf{Availability }}  & \multirow{1}{*}{\textbf{Image }}& \multirow{2}{*}{\textbf{Patients}} & \multirow{1}{*}{\textbf{Final }}& \multirow{1}{*}{\textbf{Functional }}& \multirow{1}{*}{\textbf{Articles }}\\
         
         &  &  &  \textbf{Modality}& &  \textbf{Infarct}&  \textbf{Outcome}&\textbf{Used}\\ 
         \midrule
         I-KNOW \cite{alawneh2011infarction,cheng2014influence} &  2009 & N/A  & \acrshort{mri}  & 168 & \cmark & \cmark  & 4 \\ \midrule
         DEFUSE-2 \cite{lansberg2012mri}&  2012& On Request & \acrshort{mri} & 104 & \cmark &  \cmark & 3 \\ \midrule
         % MR CLEAN Trial \cite{Berkhemer2015AStroke}&  2015& On Request & \acrshort{ncct},\acrshort{cta} & 500  & \xmark & \cmark & 5 \\ \midrule
         HERMES \cite{goyal2016endovascular}&  2016& On Request & \acrshort{ncct},\acrshort{cta} & 1287  & \xmark & \cmark & 2 \\ \midrule
         ISLES 2017 \cite{Winzeck2018ISLESMRI}&  2017& Public & \acrshort{mri} Sequences & 75  & \cmark  & \xmark & 6 \\ \midrule
         CRISP \cite{lansberg2017computed}&  2017& N/A & \acrshort{ctp} & 190 & \cmark & \cmark  & 3 \\ \midrule
         % MR CLEAN Registry \cite{compagne2022improvements}&  2018& Private & \acrshort{ncct},\acrshort{cta} & 1488* & \xmark & \cmark & 2 \\ \midrule
         iCAS \cite{thamm2019contralateral}&  2019& On Request & \acrshort{cbf} & 77 & \xmark & \cmark & 3 \\ \midrule
         ERASER \cite{fiehler2019eraser}&  2019& On Request & \acrshort{ncct},\acrshort{ctp} & 80 & \cmark & \cmark & 4 \\ \midrule
         MR CLEAN &  &  &  &   &  &  &  \\ 
          \tab Trial \cite{Berkhemer2015AStroke}&  2015& On Request & \acrshort{ncct},\acrshort{cta} & 500  & \xmark & \cmark & 5 \\ 
          \tab Registry \cite{compagne2022improvements}&  2018& Private & \acrshort{ncct},\acrshort{cta} & 1488* & \xmark & \cmark & 2 \\
         \tab NO IV \cite{lecouffe2021randomized}&  2021& On Request & \acrshort{ncct},\acrshort{cta}, \acrshort{dwi} & 539 & \xmark & \cmark & 1 \\ \midrule
         HIBISCUS-STROKE \cite{debs2021impact}&  Ongoing& On Request & \acrshort{mri}, \acrshort{ct} & 164 & \cmark & \cmark  & 2 \\ 
         
        \bottomrule
    \end{tabular}

}
\caption{Chronological overview of datasets used for \acrshort{sop}. "On request" indicates that access to associated dataset is subject to a formal application process. {\small N/A means that corresponding information was not available. * First cohort of the registry.}}
\label{table:datasets}
\end{table}

The ISLES 2017 \cite{Winzeck2018ISLESMRI} dataset comprises \gls{mri} diffusion maps (\gls{dwi}, \gls{adc}) and perfusion maps (\gls{cbv}, \gls{cbf}, \gls{mtt}, \gls{ttp}, \gls{tmax})  as well as clinical details, such as time since stroke onset, time to treatment, \gls{tici} score, and \gls{mrs} score. Additionally, ground truth labels include segmentation maps acquired from follow-up stroke imaging are also available. The training dataset consists of 43 patients, with results assessed on a separate test set of 32 stroke patients.

The Highly Effective Reperfusion Evaluated in Multiple Endovascular Stroke Trials (HERMES) \cite{goyal2016endovascular} collaboration was established to combine patient-level data from five key trials (MR CLEAN \cite{Berkhemer2015AStroke}, ESCAPE \cite{goyal2015randomized}, REVASCAT \cite{jovin2015thrombectomy}, SWIFT PRIME \cite{saver2015stent}, and EXTEND IA \cite{campbell2015endovascular}) totalling 1287 patients, admitted between late 2010 and late 2014 . The trial investigators aimed to answer outstanding questions about the effectiveness of thrombectomy in different groups of patients with large-vessel ischaemic stroke. The primary outcome of the trial was functional outcome at 90 days.

The \gls{mrclean} \cite{Berkhemer2015AStroke} dataset was collected as part of a randomised clinical trial that compared intra-arterial treatment to usual care for patients with proximal arterial occlusion in the anterior circulation, and includes data from 500 patients treated at 16 medical centres in the Netherlands. The dataset includes a baseline (\ie the first scan when the patient was admitted to the hospital) \gls{ncct} and \gls{cta} , 24-hour follow-up  \gls{ncct} and \gls{cta} scans, and additionally a 1-week follow-up \gls{ncct} scan. Additionally, \gls{mrclean} contains clinical metadata information such as patient demographics (age, gender), medical history and stroke scores (\eg \gls{nihss}), imaging biomarkers (\eg \gls{aspects}) and outcome data \gls{mrs}.

{\bfseries \gls{mrs} Score --} the \gls{mrs} score is used to assess the success rate and functional outcome of stroke treatment \cite{rankin1957cerebral,van1988interobserver} and is based on the degree of disability at 90 days after a patient received stroke treatment. It is assessed by an independent observer and scored on a standardised scale as seen in Table \ref{table:mrsscore_descript}. {Initially, \gls{mrs} was a 5-point scale ranging from 1 to 5 \cite{rankin1957cerebral}, later becoming a 6-level scale by adding 0 for patients with no symptoms \cite{van1988interobserver}, and recently \gls{mrs} has} {become} a 7-point scale that varies from 0 (no symptoms) to 6 (death of patient) \cite{quinn2009reliability}. A \gls{mrs} score between 0-2 indicates a good outcome, characterised by functional independence, which means that a patient {is able to perform daily activities without requiring assistance from another person}. This also means they were favourable to treatment and that the treatment was successful. On the other hand,  a \gls{mrs} score between 3-6 represents a bad outcome and functional dependence. 

% \vspace*{5mm}
\begin{table}[ht]
\centering
\caption{The modified Rankin Score - a basic explanation of each \acrshort{mrs} score.}
    \label{table:mrsscore_descript}

\small
\resizebox{\textwidth}{!}{
        \begin{tabular}{@{}cl@{}} 
    
         \toprule
         \textbf{Score}  & \textbf{Grade of Disability}  \\ 
          \midrule
         \small 0 &{No symptoms at all} \\ 
          \small 1 &{No significant disability despite symptoms; able to carry out all usual duties and activities} \\ 
          \small 2 &{Slight disability; unable to carry out all previous activities, but able to look after own affairs} \\ 
          \small 3 &{Moderate disability; requires some help, but able to walk without assistance} \\ 
          \small 4 &{Moderately severe disability; unable to walk and to attend to own bodily needs without assistance} \\ 
          \small 5 &{Severe disability; bedridden, incontinent and requiring constant nursing care and attention} \\ 
          \small 6 &{Dead} \\ 
     
        \bottomrule
        \end{tabular}    
}
\end{table}

{{\bfseries \acrfull{nihss} --}
the \gls{nihss} is a standardised tool used by healthcare professionals to assess the severity of stroke by way of a 15-item examination that assesses a patient's level of consciousness, speech, motor skills, vision, balance and sensory function. Each item is scored between 0 (normal) and a maximum depending on the specific function (e.g. 4 for worst motor strength), with higher total scores (0-42) indicating greater stroke severity. This score is used to guide treatment decisions, such as clot-busting medication or clot removal surgery, and to monitor a patient's recovery. Importantly, the effectiveness of the \gls{nihss} may be limited by the patient's level of cooperation. As the scale relies on the patient's ability to follow instructions and communicate, those with stroke-related impairments or pre-existing neurological conditions may not be accurately assessed on arrival at hospital or during subsequent assessments.}

{{\bfseries \acrfull{tici} --} the \gls{tici} scale provides a standardised method for grading blood vessel re-opening using digital subtraction angiography (DSA) imaging in interventional radiology following stroke treatment, particularly thrombectomy.  This scale, ranging from 0 (no flow) to 3 (complete reperfusion), allows doctors to assess the success of clot removal based on the observed blood flow patterns. Higher scores ( 2b and 3) indicate a successful procedure and a greater likelihood of positive patient outcomes.}

{{\bfseries \acrfull{aspects} --} {the \gls{aspects} score is a widely used tool for assessing the extent of early ischaemic changes in patients with \gls{ais}. This score is derived from \gls{ncct} scans and quantifies the volume of brain tissue affected by ischaemia. The score is determined by the presence or absence of ischaemic changes in specific brain regions, scales from 0 to 10. Higher score correlates with better functional outcomes, making it useful information for predicting immediate treatment response and short-term prognosis \cite{esmael2021predictive,deng2022nihss,han2024subtyping}.  However, it is important to note that the primary purpose of \gls{aspects} is to evaluate initial brain damage and inform immediate treatment decisions, rather than evaluating long-term treatment effectiveness like \gls{mrs} \cite{goyal2011effect,demeestere2018alberta}.}

%%%%%%%%%%%%%%%%%%%%%%%%%%%%%%%%%%%%%%%%%%%%%%%%%%%%%%%%%%%%%%%%%

\section{{Deep Learning for Stroke Outcome Prediction}}
\label{sec:deep_learning_stroke}

{In recent decades, a variety of techniques, from image thresholding \cite{Chawla2009AImages} to \gls{dl} \cite{ kamnitsas2017ensembles,gautam2021towards,neethi2022stroke}, have been utilised for stroke lesion detection, segmentation, and classification. However, the focus has shifted now beyond segmenting lesions to using {all available} information to specifically predict stroke treatment outcomes \cite{Bacchi2019DeepStudy, hilbert2019data,Samak_2022}.}

Previous research on predicting treatment outcomes using patient imaging {and/or} clinical data concentrates on two main areas: (1) the evolution and final appearance of stroke lesion volume and (2) the prediction of \gls{mrs} scores.
This section  reviews  and summarises key {contributions, methods, and} applications from previous studies, categorised by their {\it prediction target} {(lesion and functional outcome)} and {\it data type} {(imaging only and imaging with clinical information)} used for estimating stroke treatment outcomes. Thus next, Section \ref{sec:outcome_final_infarct} addresses studies that assess stroke treatment success by predicting the final lesion and  Section \ref{sec:outcome_mrs} then reviews studies that predict functional outcomes, \gls{mrs} scores {after stroke treatment}. 

\subsection{Final Infarct Prediction}
\label{sec:outcome_final_infarct}

\begin{table}
\renewcommand{\arraystretch}{0.77}
\centering
\large
\resizebox{\textwidth}{!}{
    \begin{tabular}{lclllccll}
        \toprule
        \multirow{2}{*}{\textbf{Study}} & 
           \multirow{2}{*}{\textbf{Year}} & 
           \multirow{1}{*}{\textbf{Patient}} &
           \multirow{1}{*}{\textbf{Treatment}} & 
           \multirow{2}{*}{\textbf{Method}} & 
           \multirow{1}{*}{\textbf{Input}} & 
           \multirow{1}{*}{\textbf{Number of}} & 
           \multirow{1}{*}{\textbf{Data}} &
           \multirow{1}{*}{\textbf{Best}} \\
          &  & \textbf{Subgroup}  & \textbf{Type}  & & \textbf{Modality }  & \textbf{Patients} & \textbf{Split} & \textbf{Result}  \\ 
        \midrule

        \multirow{2}{*}{\citet{Stier_2015}} & 
           \multirow{2}{*}{2015} & 
 
           \multirow{2}{*}{\acrshort{ais}} & 
           \multirow{2}{*}{\acrshort{evt}} & 
           \multirow{2}{*}{\acrshort{cnn}} & 
           \multirow{2}{*}{\acrshort{mri}, \acrshort{tmax}} & 
           \multirow{2}{*}{19} &
           \multirow{2}{*}{\acrshort{loocv}} &
           \multirow{2}{*}{\acrshort{auc}: 0.86} \\
          & & & &  &   & &    &  \\ 
          \midrule 

        \multirow{2}{*}{\citet{choi2016ensemble}} & 
           \multirow{2}{*}{2016} & 
  
           \multirow{2}{*}{\acrshort{ais}} & 
           \multirow{2}{*}{\acrshort{nr}} & 
           \multirow{2}{*}{U-Net} & 
           \multirow{1}{*}{\acrshort{mri}} & 
           \multirow{2}{*}{49} &
           \multirow{1}{*}{19 test} &
           \multirow{1}{*}{\acrshort{auc}: 0.74} \\
          &  & & &  & Sequences  &  & cases    & \acrshort{dice}: 0.57  \\ 
          \midrule 

        \multirow{2}{*}{\citet{Nielsen_2018}} & 
           \multirow{2}{*}{2018} & 
   
           \multirow{2}{*}{\acrshort{ais}} & 
           \multirow{2}{*}{\acrshort{rtpa}} & 
           \multirow{2}{*}{\acrshort{cnn}} & 
           \multirow{2}{*}{\acrshort{mri}} & 
           \multirow{2}{*}{222} &
           \multirow{2}{*}{85:15} &
           \multirow{2}{*}{\acrshort{auc}: 0.88} \\
          & & & &  &   & &    &  \\ 
          \midrule 

         \multirow{2}{*}{\citet{lucas2018learning}*} & 
           \multirow{2}{*}{2018} & 

           \multirow{2}{*}{\acrshort{ais}} & 
           \multirow{2}{*}{\acrshort{evt}} & 
           \multirow{1}{*}{{U-Net}} & 
           \multirow{2}{*}{\acrshort{ctp}} & 
           \multirow{2}{*}{29} &
           \multirow{1}{*}{5-fold } &
           \multirow{2}{*}{\acrshort{dice}: 0.46} \\
          & & & & + \acrshort{cae} &   & &  \acrshort{fcv}  &  \\ 
          \midrule 

         \multirow{2}{*}{\citet{pinto2018enhancing}} & 
           \multirow{2}{*}{2018} & 
   
           \multirow{2}{*}{\acrshort{ais}} & 
           \multirow{2}{*}{\acrshort{evt}} & 
           \multirow{1}{*}{multi U-Net} & 
           \multirow{1}{*}{\acrshort{dwi}} & 
           \multirow{2}{*}{75} &
           \multirow{1}{*}{32 test} &
           \multirow{2}{*}{\acrshort{dice}: 0.29} \\
          &  & & & + GRU & \acrshort{pwi}  & &   cases &  \\ 
          \midrule 

         \multirow{2}{*}{\citet{ho2019predicting}} & 
           \multirow{2}{*}{2019} & 
     
           \multirow{2}{*}{\acrshort{mca}} & 
           \multirow{2}{*}{\acrshort{evt}} & 
           \multirow{2}{*}{\acrshort{cnn}} & 
           \multirow{1}{*}{\acrshort{mri}, \acrshort{pwi}} & 
           \multirow{2}{*}{48} &
           \multirow{1}{*}{10-fold } &
           \multirow{2}{*}{\acrshort{auc}: 0.871} \\
          &  & & &  &  \acrshort{flair} & &   \acrshort{fcv} &  \\ 
          \midrule 

         \multirow{2}{*}{\citet{Debs_2019}} & 
           \multirow{2}{*}{2019} & 
    
           \multirow{1}{*}{Anterior} & 
           \multirow{2}{*}{\acrshort{tpa}} & 
           \multirow{2}{*}{\acrshort{cnn}} & 
           \multirow{1}{*}{Synthetic} & 
           \multirow{1}{*}{100} &
           \multirow{1}{*}{8 test } &
           \multirow{2}{*}{\acrshort{dice}: 0.45} \\
          & &circulation & &  & \acrshort{pwi}  &simulations &    cases&  \\ 
          \midrule 

        \multirow{2}{*}{\citet{Sales_Barros_2019}} & 
           \multirow{2}{*}{2019} & 
 
           \multirow{2}{*}{\acrshort{ais}} & 
           \multirow{2}{*}{\acrshort{evt}} & 
           \multirow{2}{*}{\acrshort{cnn}} & 
           \multirow{2}{*}{\acrshort{ct}} & 
           \multirow{2}{*}{1026} &
           \multirow{1}{*}{396 test } &
           \multirow{1}{*}{\acrshort{icc}: 0.88} \\
          &  & & &  &   & & cases   &  \acrshort{dice}: 0.57 \\ 
          \midrule 

         \multirow{2}{*}{\citet{Debs_2020}} & 
           \multirow{2}{*}{2020} & 
       
           \multirow{1}{*}{Anterior} & 
           \multirow{2}{*}{\acrshort{tpa}} & 
           \multirow{2}{*}{\acrshort{cnn}} & 
           \multirow{2}{*}{\acrshort{pwi}} & 
           \multirow{1}{*}{125000} &
           \multirow{1}{*}{8 test} &
           \multirow{2}{*}{\acrshort{dice}: 0.40} \\
          &  & circulation& &  &   &patches & cases   &  \\ 
          \midrule 

         \multirow{2}{*}{\citet{yu2020use}} & 
           \multirow{2}{*}{2020} & 
  
           \multirow{1}{*}{{Anterior}} & 
           \multirow{2}{*}{\acrshort{evt}} & 
           \multirow{2}{*}{\acrshort{cnn}} & 
           \multirow{1}{*}{\acrshort{mri}} & 
           \multirow{2}{*}{182} &
           \multirow{1}{*}{5-fold} &
           \multirow{1}{*}{\acrshort{auc}: 0.92} \\
          &  & circulation & &  & Sequences  & & \acrshort{fcv}   & \acrshort{dice}: 0.53  \\ 
          \midrule 

         \multirow{2}{*}{\citet{Lucas_2020}} & 
           \multirow{2}{*}{2020} & 
          
           \multirow{2}{*}{\acrshort{ais}} & 
           \multirow{2}{*}{\acrshort{evt}} & 
           \multirow{1}{*}{\acrshort{cnn} } & 
           \multirow{2}{*}{\acrshort{ctp}} & 
           \multirow{2}{*}{29} &
           \multirow{1}{*}{5-fold } &
           \multirow{2}{*}{{F1}: 0.74} \\
          &  & & & + \acrshort{pca} &   & & \acrshort{fcv}   &  \\ 
          \midrule 

         \multirow{2}{*}{\citet{robben2020prediction}} & 
           \multirow{2}{*}{2020} & 
   
           \multirow{2}{*}{\acrshort{lvo}} & 
           \multirow{2}{*}{\acrshort{evt}} & 
           \multirow{1}{*}{\acrshort{cnn}} & 
           \multirow{2}{*}{\acrshort{ctp}} & 
           \multirow{2}{*}{188} &
           \multirow{1}{*}{5-fold } &
           \multirow{2}{*}{\acrshort{dice}: 0.48} \\
          &  &  & & DeepMedic &   & &   \acrshort{fcv} &  \\ 
          \midrule 

         \multirow{2}{*}{\citet{Pinto2021prediction}} & 
           \multirow{2}{*}{2021} & 
 
           \multirow{2}{*}{\acrshort{ais}} & 
           \multirow{2}{*}{\acrshort{evt}} & 
           \multirow{1}{*}{{End2End}} & 
           \multirow{2}{*}{DSC-\acrshort{mri}} & 
           \multirow{2}{*}{75} &
           \multirow{1}{*}{32 test } &
           \multirow{2}{*}{\acrshort{dice}: 0.31} \\
          & & & & \acrshort{cnn} &   & &  cases  &  \\ 
          \midrule 

         \multirow{2}{*}{\citet{debs2021impact}} & 
           \multirow{2}{*}{2021} & 
  
           \multirow{2}{*}{\acrshort{lvo}} & 
           \multirow{2}{*}{\acrshort{evt}} & 
           \multirow{1}{*}{Multi-scale} & 
           \multirow{1}{*}{\acrshort{dwi}} & 
           \multirow{2}{*}{109} &
           \multirow{1}{*}{5-fold} &
           \multirow{1}{*}{\acrshort{auc}: 0.87} \\
          & & & & U-Net &  \acrshort{pwi}  & &  \acrshort{fcv}  & \acrshort{dice}:0.44  \\ 
          \midrule 

         \multirow{2}{*}{\citet{Hokkinen_2021}} & 
           \multirow{2}{*}{2021} & 

           \multirow{2}{*}{\acrshort{lvo}} & 
           \multirow{2}{*}{\acrshort{evt}} & 
           \multirow{2}{*}{\acrshort{cnn}} & 
           \multirow{2}{*}{\acrshort{cta}} & 
           \multirow{2}{*}{89} &
           \multirow{1}{*}{50 test } &
           \multirow{2}{*}{\acrshort{dice}: 0.60} \\
          &  & & &  &   & & cases &  \\ 
          \midrule 

         \multirow{2}{*}{\citet{pinto2021combining}*} & 
           \multirow{2}{*}{2021} & 
      
           \multirow{2}{*}{\acrshort{ais}} & 
           \multirow{2}{*}{\acrshort{evt}} & 
           \multirow{1}{*}{\acrshort{rbm}} & 
           \multirow{1}{*}{\acrshort{mri}} & 
           \multirow{2}{*}{75} &
           \multirow{1}{*}{32 test } &
           \multirow{2}{*}{\acrshort{dice}: 0.38} \\
          & & & & \acrshort{cnn} & Sequences  & & cases &  \\ 
          \midrule 

         \multirow{2}{*}{\citet{Chen_2021}} & 
           \multirow{2}{*}{2021} & 
           
           \multirow{2}{*}{\acrshort{ais},} & 
           \multirow{2}{*}{{\acrshort{tpa}}} & 
           \multirow{1}{*}{Ensemble} & 
           \multirow{2}{*}{\acrshort{dwi}} & 
           \multirow{2}{*}{40} &
           \multirow{2}{*}{70:30} &
           \multirow{2}{*}{\acrshort{auc}: 0.898} \\
          & &  & &U-Net + \acrshort{ml}  &   & &    &  \\ 
          \midrule 

         \multirow{2}{*}{\citet{amador2022predicting}} & 
           \multirow{2}{*}{2022} & 
    
           \multirow{2}{*}{\acrshort{ais}} & 
           \multirow{2}{*}{\acrshort{evt}} & 
           \multirow{1}{*}{{U-Net}} & 
           \multirow{2}{*}{4D \acrshort{ctp}} & 
           \multirow{2}{*}{147} &
           \multirow{1}{*}{10-fold } &
           \multirow{2}{*}{\acrshort{dice}: 0.45} \\
          &  & & & Temporal-\acrshort{cnn}  &   & &  \acrshort{fcv}  &  \\ 
          \midrule 

        \multirow{2}{*}{\citet{He_2022}} & 
           \multirow{2}{*}{2022} & 

           \multirow{2}{*}{\acrshort{lvo}} & 
           \multirow{2}{*}{\acrshort{evt}} & 
           \multirow{2}{*}{{U-Net}} & 
           \multirow{2}{*}{\acrshort{ctp}} & 
           \multirow{2}{*}{110} &
           \multirow{2}{*}{60:20:20 } &
           \multirow{1}{*}{\acrshort{auc}: 0.93} \\
          &  & & &  &   & &    &\acrshort{dice}:0.67  \\ 
          \midrule 

         \multirow{2}{*}{\citet{Winder_2022}} & 
           \multirow{2}{*}{2022} & 
     
           \multirow{2}{*}{\acrshort{lvo}} & 
           \multirow{1}{*}{\acrshort{evt}} & 
           \multirow{2}{*}{{U-Net}} & 
           \multirow{2}{*}{\acrshort{ctp}} & 
           \multirow{2}{*}{145} &
           \multirow{2}{*}{68:12:20 } &
           \multirow{2}{*}{\acrshort{dice}: 0.289} \\
          & & & \acrshort{tpa}&  &   & &    &  \\ 
          \midrule 

         \multirow{2}{*}{\citet{amador2022hybrid}*} & 
           \multirow{2}{*}{2022} & 

           \multirow{2}{*}{\acrshort{ais}} & 
           \multirow{2}{*}{\acrshort{evt}} & 
           \multirow{1}{*}{\acrshort{cnn} +} & 
           \multirow{2}{*}{\acrshort{ctp}} & 
           \multirow{2}{*}{45} &
           \multirow{1}{*}{10-fold } &
           \multirow{2}{*}{\acrshort{dice}: 0.45} \\
          & & & & Transformer &   & &  \acrshort{fcv}  &  \\ 
          \midrule 

         \multirow{2}{*}{\citet{Wouters_2022}} & 
           \multirow{2}{*}{2022} & 
      
           \multirow{2}{*}{\acrshort{lvo}} & 
           \multirow{2}{*}{\acrshort{evt}} & 
           \multirow{1}{*}{\acrshort{cnn}} & 
           \multirow{2}{*}{\acrshort{ctp}} & 
           \multirow{2}{*}{127} &
           \multirow{1}{*}{5-fold \acrshort{fcv},} &
           \multirow{2}{*}{\acrshort{auc}: 0.88} \\
          & & & & DeepMedic &   &  &101 ext. val&  \\ 
          \midrule 

         \multirow{2}{*}{\citet{Lee_2022}*} & 
           \multirow{2}{*}{2022} & 

           \multirow{2}{*}{\acrshort{ais}} & 
           \multirow{2}{*}{\acrshort{tpa}} & 
           \multirow{2}{*}{{U-Net}} & 
           \multirow{1}{*}{\acrshort{dwi}} & 
           \multirow{2}{*}{472} &
           \multirow{1}{*}{5-fold \acrshort{fcv}} &
           \multirow{2}{*}{\acrshort{dice}: 0.486} \\
          & & & &  & \acrshort{pwi}  & &55 ext. test&  \\ 
          \midrule 

         \multirow{2}{*}{Nazari \etal\cite{Nazari_Farsani_2023}} & 
           \multirow{2}{*}{2023} & 

           \multirow{2}{*}{\acrshort{ais}} & 
           \multirow{1}{*}{\acrshort{evt}} & 
           \multirow{1}{*}{{Attention-gate}} & 
           \multirow{1}{*}{\acrshort{dwi}} & 
           \multirow{2}{*}{445} &
           \multirow{1}{*}{5-fold} &
           \multirow{1}{*}{\acrshort{auc}: 0.91} \\
          & & &\acrshort{tpa} & U-Net &\acrshort{adc} & &   \acrshort{fcv} & \acrshort{dice}:0.50  \\ 
        \midrule 

         \multirow{2}{*}{\citet{Amador_2024}} & 
           \multirow{2}{*}{2024} & 
           \multirow{2}{*}{\acrshort{ais}} & 
           \multirow{2}{*}{\acrshort{evt}} & 
           \multirow{1}{*}{\acrshort{cnn}} & 
           \multirow{2}{*}{\acrshort{ctp}} & 
           \multirow{2}{*}{70} &
           \multirow{2}{*}{10-fold \acrshort{fcv}} &
           \multirow{2}{*}{\acrshort{dice}: 0.26} \\
          & & & & CrossAttention & & &    &   \\ 
        \midrule 

         \multirow{2}{*}{\citet{Palsson_2024}} & 
           \multirow{2}{*}{2024} & 
           \multirow{2}{*}{\acrshort{ais}} & 
           \multirow{2}{*}{\acrshort{evt}} & 
           \multirow{2}{*}{U-Net} & 
           \multirow{1}{*}{\acrshort{ncct}} & 
           \multirow{2}{*}{404} &
           \multirow{1}{*}{80:20} &
           \multirow{2}{*}{\acrshort{dice}: 0.37} \\
          & & & &  &\acrshort{cta} & & 5-fold \acrshort{fcv}  &   \\ 
        \midrule 

         \multirow{2}{*}{\citet{Gutierrez_2024}} & 
           \multirow{2}{*}{2024} & 
           \multirow{2}{*}{\acrshort{ais}} & 
           \multirow{1}{*}{\acrshort{evt}} & 
           \multirow{2}{*}{\acrshort{cgan}} & 
           \multirow{1}{*}{\acrshort{ncct}} & 
           \multirow{2}{*}{147} &
           \multirow{2}{*}{5-fold \acrshort{fcv}} &
           \multirow{2}{*}{\acrshort{dice}: 0.37} \\
          & & & \acrshort{tpa} &  &\acrshort{ctp} & &   &   \\ 
    
        \bottomrule
    \end{tabular}

}
\caption{Overview of the studies that perform final infarct prediction as stroke treatment outcome.  {\tiny Abbreviations are: {\acrfull{is}, \acrfull{ais}, \acrfull{mca}, \acrfull{lvo},  \acrfull{evt}, \acrfull{rtpa}, \acrfull{tpa}, \acrfull{cnn}, \acrfull{cae}, \acrfull{cgan}, \acrfull{pca}, \acrfull{rbm}, \acrfull{mri},
\acrfull{tmax}, \acrfull{ctp}, \acrfull{dwi}, \acrfull{pwi}, \acrfull{flair}, \acrfull{ct}, \acrfull{ncct}, \acrfull{cta}, \acrfull{adc}, \acrfull{loocv}, \acrfull{fcv}, \acrfull{auc}, \acrfull{dice}, \acrfull{icc}, \acrfull{nr}}. {* indicates that the code used in the study is available.} }}
\label{table:outcome_lesion}
\end{table}

This section focuses on predicting stroke treatment outcomes by estimating the final appearance of the stroke lesion on follow-up scans. {While follow-up scans can be acquired within a broad time frame ranging from 1 hour to 90 days \cite{gaudinski2008establishing,pinto2018enhancing,Wang2022}, most studies generally adopt a time point of one week (5-7 days) \cite{krongold2015final,yu2020use,Lee_2022,Liu_2023,Nazari_Farsani_2023}.} Researchers {have mainly addressed} this task by developing models trained on datasets of expert-annotated follow-up scans. These models analyse a patient's initial scan and predict the final lesion segmentation map, which is essentially a visual representation of the extent of the stroke. This information provides physicians with a valuable tool to develop more effective treatment plans. Throughout this article, the terms "final infarct", "tissue fate", "follow-up lesion" and "final lesion" are used interchangeably {to mean the same outcome and to remain consistent with terms used in the community.}

{An overview of various studies examined for final infarct prediction is presented in Table \ref{table:outcome_lesion}.} Advanced imaging techniques, such as \gls{dwi}, \gls{pwi}, and \gls{ctp} provide valuable information to assess tissue viability and predict stroke outcomes. \gls{dl} models have emerged as powerful tools in this domain, leveraging the ability to extract complex patterns and relationships from high-dimensional imaging data. Studies in this area have mainly used U-Net-like \gls{cnn} network architectures {and incorporated different training strategies and modules such as} multi-scale input \cite{pinto2018enhancing,debs2021impact}, ensemble learning \cite{choi2016ensemble,Chen_2021}, transformer \cite{amador2022hybrid} and attention modules \cite{Nazari_Farsani_2023,Amador_2024}.

\citet{Stier_2015} presented a deep learning model trained on randomly sampled local patches extracted from \gls{tmax} feature maps of \gls{mri} scans acquired immediately after symptom onset. Their approach outperformed traditional single-voxel regression models, such as \cite{nguyen2008stroke}, in predicting tissue survival outcomes in acute ischaemic stroke patients. Similarly, \citet{Nielsen_2018} developed {a generalised linear model, a shallow \gls{cnn} and a deep \gls{cnn} based on the SegNet \cite{badrinarayanan2017segnet}} model, trained on \gls{mri}-based perfusion maps of 222 patients. {Their results demonstrated that the deep \gls{cnn} model outperformed other models in predicting final infarct volumes}. 

%MULTIMODAL
The integration of various data sources shows great potential for improving the performance of \gls{dl} models. \citet{pinto2018enhancing} provided an example of this approach by proposing a method that used \gls{mri} maps and clinical data to predict the outcome of stroke lesions at a 3-month follow-up. The model achieved improved accuracy by incorporating clinical information at both the population and individual patient levels. {\citet{Amador_2024} developed a novel \gls{dl} model using cross-attention to incorporate the spatio-temporal nature of 4D \gls{ctp} imaging and clinical variables to predict final infarct. Although {their multimodal} model performed slightly worse, with a \gls{dice} score of 0.25, compared to {their} unimodal model \gls{dice} score of 0.26, it significantly outperformed the unimodal approach in estimating lesion volume (mean error: 19 ml vs. 35 ml). In addition, the model generated attention maps and valuable information on how clinical variables influence stroke outcomes at individual patient level.}

\begin{figure}[t] %[ht]
    \centering
    \includegraphics[width=.85\linewidth]{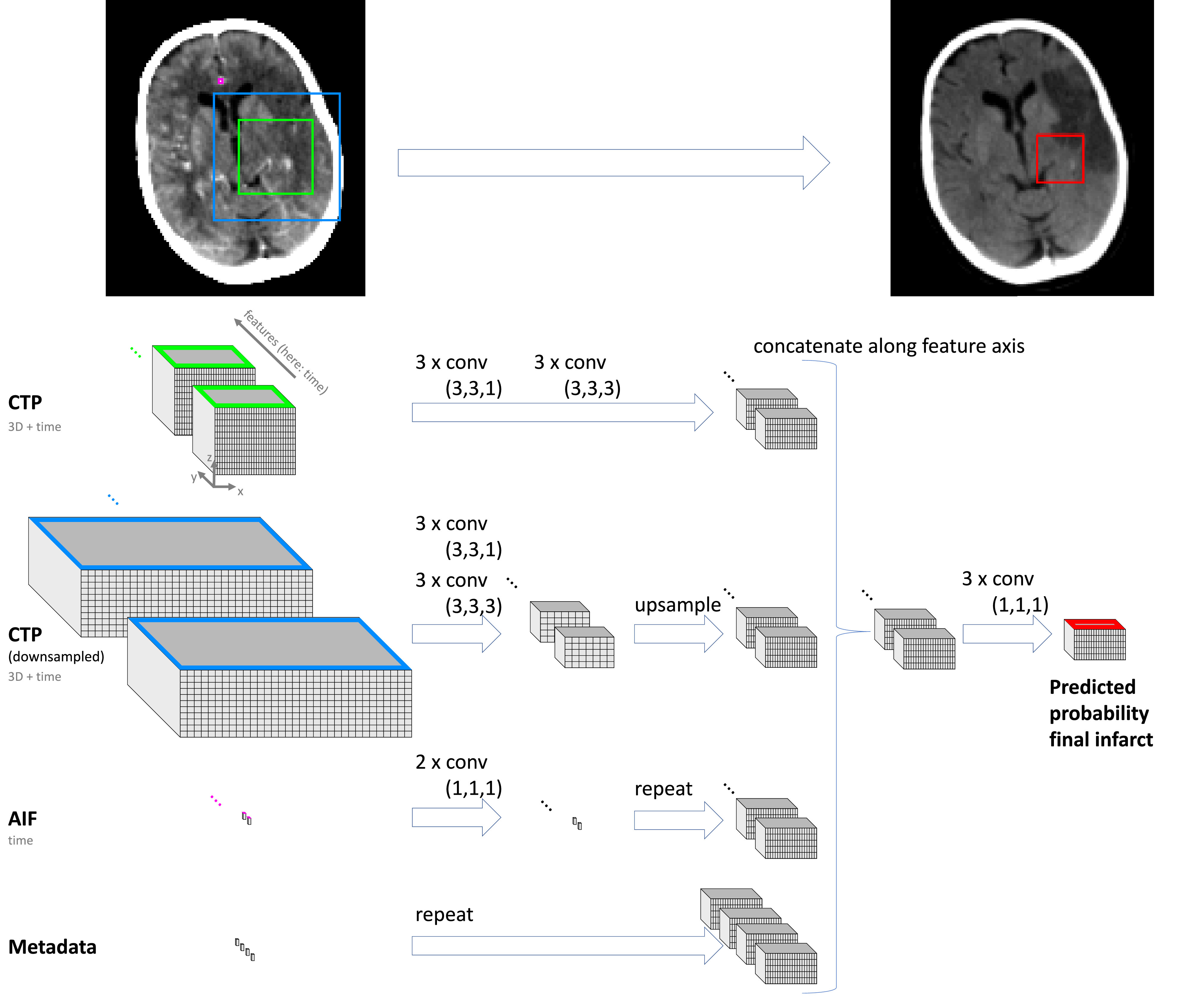}
    \caption{The multimodal architecture introduced in \cite{robben2020prediction}. The network exploits imaging and clinical information to predict the follow-up stroke lesion. {\small \gls{ctp}: computed tomography perfusion, AIF: arterial input function which refers to the measure of blood concentration that enters the affected brain tissue after a stroke, Metadata: clinical information \eg time to scanning, recanalization, completeness of recanalization. Yellow and blue rectangles represent patches extracted from \gls{ctp} scan, purple rectangle represents AIF in the volume of interest and red rectangle is the target region of interest to segment lesion on follow-up \gls{ncct} scan.} Figure from \cite{robben2020prediction}.}
    \label{figc2:robben_network}
\end{figure}

\citet{robben2020prediction} developed a \gls{cnn} model, based on DeepMedic \cite{kamnitsas2017a}, using spatio-temporal \gls{ctp} data and clinical information to predict follow-up infarct lesions, as shown in Fig. \ref{figc2:robben_network}. The model architecture comprises four branches  that extract the features of \gls{ctp} images {and their downsampled versions}, arterial input function, and metadata. These features are concatenated and fed through three convolutional layers to generate the final infarct segmentation map.  \citet{debs2021impact}  investigated \gls{cnn}s to predict stroke infarct volume, incorporating reperfusion status (successful or failed blood flow restoration) into the model.  Their findings suggest that \glspl{cnn} outperform the current clinical standard, the perfusion-diffusion mismatch model \cite{davalos2004clinical}, in predicting infarct volume for both reperfused and non-reperfused patients (reperfused patients: \gls{auc} = 0.87 vs. 0.79; non-reperfused patients: \gls{auc} = 0.81  vs. 0.73 , in \gls{cnn} vs. perfusion–diffusion mismatch models, respectively). \citet{Pinto2021prediction} demonstrated the effectiveness of integrating information from perfusion dynamic susceptibility \gls{mri} (DSC-MRI) and parametric maps in predicting ischaemic stroke tissue outcomes. Their end-to-end architecture, which is based on U-Net and fully \gls{cnn}, extracted features from raw perfusion DSC-MRI to complement standard parametric maps, achieving competitive results in ISLES 2017 competition \cite{Winzeck2018ISLESMRI} and reaching the second highest average \gls{dice} of 0.31.

To improve model performance, researchers have explored ensemble approaches that integrate several deep neural networks {or different configurations within the same architecture}.  A key example was presented in \cite{choi2016ensemble}, where an ensemble model {of various configurations of a U-Net architecture, including different patch sizes (multi-scale image patches), number of patches, and number of convolutional filters} with a multi-step training strategy was employed to overcome challenges of class imbalance {due to lesion size compared to whole brain}, and limited data {of} 30 training cases. This approach, utilised multi-scale image patches, and achieved outstanding performance in predicting both lesion outcome and clinical outcomes, ranking amongst the leaders in the ISLES 2016 Challenge \cite{Winzeck2018ISLESMRI}. {Similarly, \citet{Chen_2021} introduced AUNet, an ensemble model that integrated \gls{dl} and \gls{ml} techniques.  AUNet brought together an adaptive linear ensemble model (ALEM), incorporating \gls{rf}, extremely randomised trees, and \gls{xgb}, with a U-Net \cite{ronneberger2015u} to capture voxel-wise information and learn spatial relationships. Their method effectively predicted infarct volume in patients with acute ischaemic stroke under varying recanalization conditions, achieving an \gls{auc} of 0.898 on a dataset of 40 patients.}

\begin{figure}[!t]
    \includegraphics[width=\linewidth]{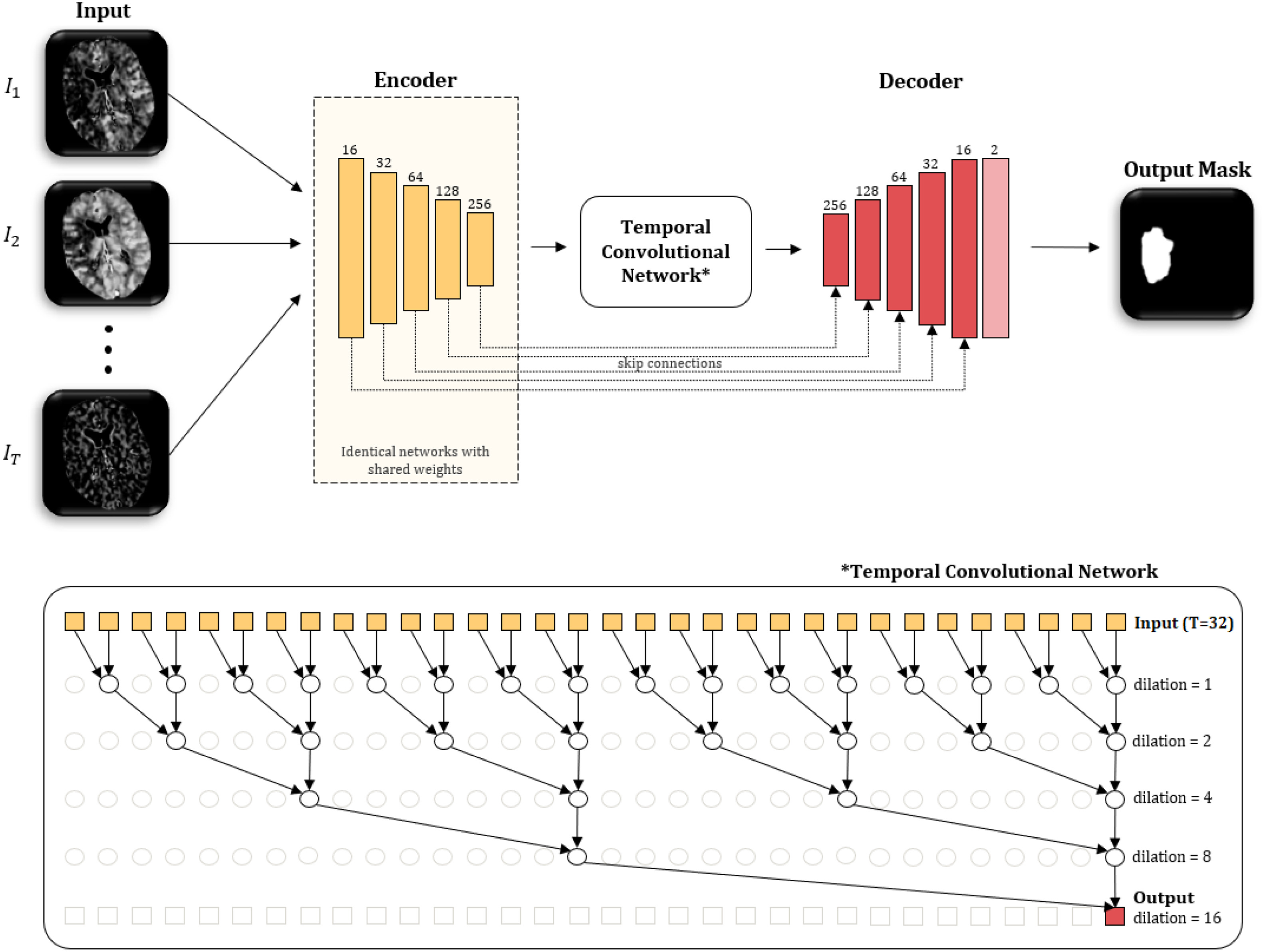}
    \caption{The overview of the temporal U-Net proposed by \cite{amador2022predicting} to estimate the final lesion mask of stroke. {Each \gls{ctp} scan is fed separately into an encoder, and the temporal convolutional block encodes temporal information and pass the features to the decoder for the purpose of predicting the final lesion mask.} Figure from \cite{amador2022predicting}.}
    \label{figc2:amador_network}
\end{figure}

Recent research shows that \gls{dl} models can directly analyse raw 4D \gls{ctp} data to predict stroke lesion outcomes, removing post-processing steps to generate 3D perfusion parameter maps and potentially capturing spatio-temporal information more effectively \cite{amador2022predicting, amador2022hybrid}. \citet{amador2022predicting} demonstrated the benefit of using spatio-temporal information of raw 4D \gls{ctp} by employing a U-Net-like architecture (see Fig. \ref{figc2:amador_network}) with shared-weight encoders for spatial feature extraction, followed by a temporal convolutional network (TCN) to integrate temporal information. A decoder then generated a segmentation map which was then post-processed for refinement. Their 3D+time model, validated through a 10-fold cross-validation on 147 patients {curated from ERASER  \cite{fiehler2019eraser} and custom data}, obtained a notably higher mean \gls{dice} of 0.45 compared to both their 2D+time version (0.43 \gls{dice}) and the state-of-the-art method \cite{Nielsen_2018} based on perfusion maps (0.38 \gls{dice}).

\citet{Gutierrez_2024} introduced a annotation-free method that directly predicted follow-up \gls{ncct} images of acute ischaemic stroke patients from 4D \gls{ctp} scans, bypassing the need for complex perfusion analysis or manual annotation of segmentation maps. Their proposed method first employed a temporal autoencoder network to extract features from \gls{ctp}, followed by a \gls{cgan} that utilised these features to directly predict follow-up \gls{ncct} images. The model was trained and tested on the same dataset of 147 patients as in \cite{amador2022predicting}, with separate models for thrombolysis (45 patients) and thrombectomy (102 patients) treatments. Their findings showed that their approach produced realistic follow-up image predictions (also see Fig. \ref{figc2:gutierez_network}) with a performance comparable to the {previous} state-of-the-art method \cite{amador2022predicting} which relied on manual annotation. Specifically, the 3D-TCN method in \citet{amador2022predicting} achieved \gls{dice} scores of 0.396 and 0.214 for thrombolysis and thrombectomy, respectively, while \citet{Gutierrez_2024}'s method achieved \gls{dice} scores of 0.375 and 0.214.

\begin{figure}[!t]
    \includegraphics[width=\linewidth]{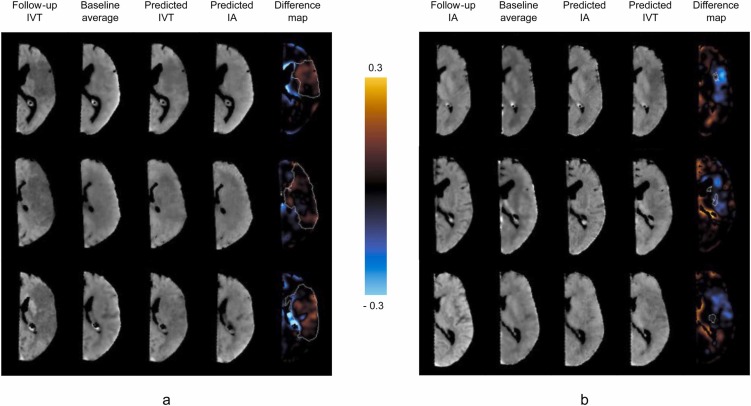}
    \caption{Sample results of \cite{Gutierrez_2024} for three patients from each treatment group of (a) thrombolysis (IVT)  and (b) thrombectomy (IA). Five columns for each patient: (1) the original follow-up \gls{ct} image, (2) the baseline image average, (3) the \gls{cgan} prediction for the patient's treatment group, (4) the \gls{cgan} prediction from the alternative treatment model, and (5) a difference map highlighting changes in intensity between predictions. The ground truth lesion mask is outlined in white within the difference map, with orange indicating increased intensity in the alternative treatment prediction (column 4) and blue indicating decreased intensity. Figure from \cite{Gutierrez_2024}.}
    \label{figc2:gutierez_network}
\end{figure}

\begin{figure}[!b] %[ht]
    \centering
    \includegraphics[width=\linewidth]{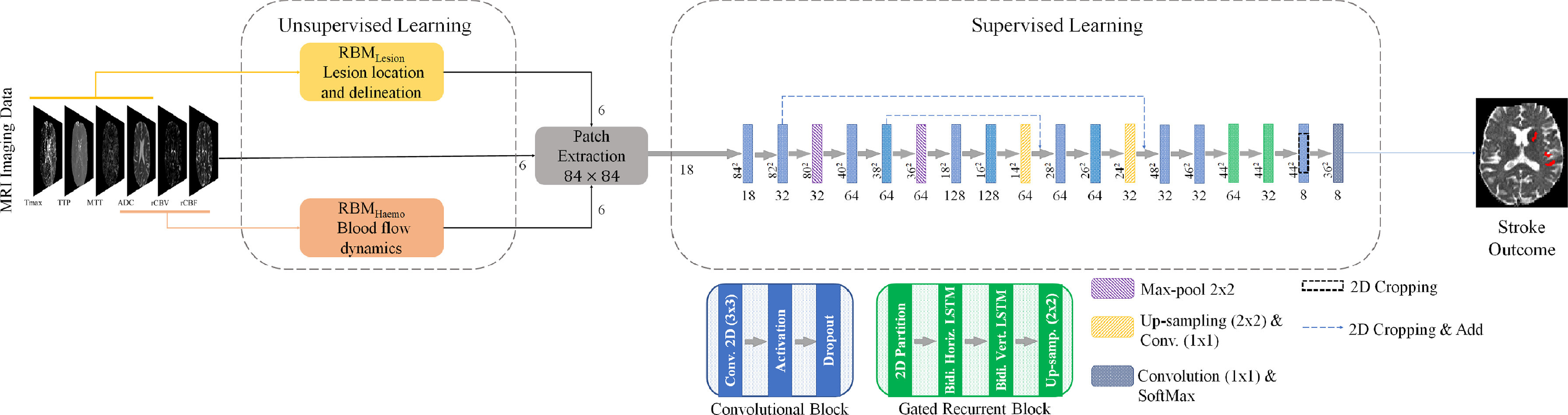}
    \caption{The architecture proposed by \cite{pinto2021combining} to predict stroke outcome. The first part of the network includes unsupervised training that uses two \acrfull{rbm} and the second part of the network comprises of  \gls{cnn} and gated-\gls{rnn}s for supervised training. Figure from \cite{pinto2021combining}.}
    \label{figc2:pinto_network}
\end{figure}

Some recent studies explored different techniques to improve model performance. For example, \citet{yu2020use} integrated an attention gate into a U-Net network \cite{oktay2018attention} to predict tissue outcome from \gls{mri} perfusion maps and obtained \gls{auc} of 0.92 and \gls{dice} of 0.53 {on a dataset consisting of iCAS \cite{zaharchuk2015abstract} and DEFUSE-2 \cite{lansberg2012mri} data}. \citet{pinto2021combining} achieved the best performance with \gls{dice} of 0.38 in the ISLES 2017 \cite{Winzeck2018ISLESMRI} dataset by employing a two-stage approach (see Fig. \ref{figc2:pinto_network}), initially using \glspl{rbm} in an unsupervised manner to identify lesion locations and blood flow characteristics, followed by a supervised stage where a \gls{cnn} based on U-Net \cite{ronneberger2015u}  and gated-\glspl{rnn} were employed to extract spatial relationships and predict stroke lesions on follow-up scans.

Synthetic data has also emerged as a valuable tool for training models, especially when real-world data is scarce. Researchers have shown the potential of synthetic data in this context by using physically realistic simulated perfusion \gls{mri} data to train models for predicting final infarct lesions \cite{Debs_2019}.  Furthermore, incorporating patient-specific details, such as arterial input functions, has demonstrably improved the accuracy of these predictions \cite{Debs_2020}. Generating synthesised data effectively addresses data limitations to potentially better  integrate domain knowledge into \gls{dl} models.

\subsection{Functional Outcome (mRS) Prediction}
\label{sec:outcome_mrs}

The studies in this category aim to assess the functional abilities of stroke patients, following intervention, by analysing unimodal and multimodal  data to estimate their \gls{mrs} scores. These  {learn} to establish the stroke treatment outcome as \textit{dichotomous} (\gls{mrs} scores 0-2 as favourable and \gls{mrs} scores 3-6 as unfavourable) or as \textit{individual} \gls{mrs} scores (0-6). At the inference stage, the model predicts the functional outcome for a new patient using imaging and/or clinical data. Consequently, these models can help clinicians determine the most appropriate treatment.

\subsubsection{Image Data Analysis}
\label{sec:outcome_image}

Predicting the functional outcome after stroke treatment is essential for guiding treatment decisions and patient management. Neuroimaging biomarkers and clinical information have commonly been used, but recent advances in \gls{dl} have introduced new opportunities to use imaging data. When clinical health records are not accessible during hospital admission, the use of imaging data offers a viable alternative to evaluate stroke outcomes.

% \eg \gls{mri}, \gls{cta} and \gls{ncct},
Table \ref{table:outcome_mrs_image} shows the studies that leverage only imaging information to predict the functional outcome of stroke treatment. These studies employ various \gls{dl} techniques, including \gls{cnn} \cite{nishi2020deep,Xia_2023}, Siamese networks \cite{Osama2020PredictingNetwork}, and attention mechanisms \cite{Moulton_2022}. The models are trained on \gls{cta} \cite{hilbert2019data}, \gls{ncct} \cite{Fang_2022,Xia_2023}, \gls{mri} \cite{Lai_2022} (including multi-parametric \gls{mri} \cite{Osama2020PredictingNetwork}),  and diffusion-weighted imaging \cite{nishi2020deep,Tolhuisen_2022} modalities.

\citet{hilbert2019data} used a ResNet model with \glspl{rfnn} \cite{jacobsen2016structured} which was trained on 2D images, to extract high-level features from a projection of the maximum intensity of a 3D \gls{cta} volume of the patient. Their model outperformed traditional biomarkers such as \gls{aspects} and ischaemic core volume to predict reperfusion (\gls{tici}) (\gls{auc} 0.65) and functional outcome (\gls{auc} 0.71) {on the \gls{mrclean} registry \cite{compagne2022improvements} dataset}. In addition, they used \gls{gradcam} \cite{selvaraju2017grad} to interpret their model result. \gls{gradcam} demonstrates the convolutional feature maps that contribute the most to prediction in input space. \citet{nishi2020deep} combined  lesion segmentation and \gls{mrs} score prediction tasks in a single U-Net model trained on \gls{dwi} as seen in Fig. \ref{figc2:nishi_network}. Using the features at the U-Net bottleneck, they performed the classification task of \gls{mrs} scores, and the output map of the U-Net decoder was used for the segmentation of the infarct core. Their model  predicted functional outcome at 90 days, outperforming standard biomarkers like the \gls{aspects} and ischaemic core volume by achieving an \gls{auc} of 0.81 in a derivation cohort of 250 patients and \gls{auc} of 0.73 in a validation cohort of 74 patients. Their dataset was collected from four hospitals in Japan and is available on request from the authors.

\begin{table}[ht]
\centering
\large
\resizebox{\textwidth}{!}{
    \begin{tabular}{lclllccll}
        \toprule
        % \specialrule{.2em}{.1em}{.1em}

        \multirow{2}{*}{\textbf{Study}} & 
           \multirow{2}{*}{\textbf{Year}} & 

           \multirow{1}{*}{\textbf{Patient}} &
           \multirow{1}{*}{\textbf{Treatment}} & 
           \multirow{2}{*}{\textbf{Method}} & 
           \multirow{1}{*}{\textbf{Input}} & 
           \multirow{1}{*}{\textbf{Number of}} & 
           \multirow{1}{*}{\textbf{Data}} &
           \multirow{1}{*}{\textbf{Best}} \\
          &  & \textbf{Subgroup}  & \textbf{Type}  & & \textbf{Modality }  & \textbf{Patients} & \textbf{Split} & \textbf{Result}  \\ 
        \midrule

        \multirow{2}{*}{\citet{hilbert2019data}} & 
           \multirow{2}{*}{2019} & 
 
           \multirow{2}{*}{\acrshort{ais}} & 
           \multirow{2}{*}{\acrshort{evt}} & 
           \multirow{1}{*}{{ResNet}} & 
           \multirow{2}{*}{\acrshort{cta}} & 
           \multirow{2}{*}{1301} &
           \multirow{1}{*}{4-fold } &
           \multirow{2}{*}{\acrshort{auc}: 0.71} \\
          & & &  & + RFNN &   & &\acrshort{fcv} &  \\ 
          \midrule 

        \multirow{2}{*}{\citet{nishi2020deep}} & 
           \multirow{2}{*}{2020} & 
 
           \multirow{2}{*}{\acrshort{lvo}} & 
           \multirow{2}{*}{\acrshort{evt}} & 
           \multirow{1}{*}{U-Net} & 
           \multirow{2}{*}{\acrshort{dwi}} & 
           \multirow{2}{*}{250} &
           \multirow{1}{*}{5-fold \acrshort{fcv} } &
           \multirow{2}{*}{\acrshort{auc}: 0.81} \\
          &  & & & + \acrshort{fc}  &   & &  74 ext. val.  &  \\ 
          \midrule 

        \multirow{2}{*}{\citet{Osama2020PredictingNetwork}} & 
           \multirow{2}{*}{2020} & 
    
           \multirow{2}{*}{\acrshort{ais}} & 
           \multirow{2}{*}{\acrshort{evt}} & 
           \multirow{1}{*}{{Siamese}} & 
           \multirow{2}{*}{\acrshort{mpmri}} & 
           \multirow{2}{*}{43} &
           \multirow{2}{*}{70:30 } &
           \multirow{2}{*}{\acrshort{acc}: 0.67} \\
           & & & & Network &   & &    &  \\ 
          \midrule 

        \multirow{2}{*}{\citet{Lai_2022}} & 
           \multirow{2}{*}{2022} & 
    
           \multirow{2}{*}{\acrshort{nr}} & 
           \multirow{2}{*}{\acrshort{nr}} & 
           \multirow{1}{*}{\acrshort{cnn}} & 
           \multirow{2}{*}{\acrshort{mri}} & 
           \multirow{2}{*}{{44}} &
           \multirow{2}{*}{90:10 } &
           \multirow{2}{*}{\acrshort{acc}: 0.932} \\
          & & & & {VGG16}  &   & &    &  \\ 
          \midrule 

        \multirow{2}{*}{\citet{Fang_2022}} & 
           \multirow{2}{*}{2022} & 
      
           \multirow{1}{*}{{Posterior}} & 
           \multirow{1}{*}{\acrshort{evt}} & 
           \multirow{1}{*}{ResNet-18} & 
           \multirow{2}{*}{\acrshort{ncct}} & 
           \multirow{2}{*}{31} &
           \multirow{1}{*}{5-fold } &
           \multirow{2}{*}{\acrshort{auc}: 0.74} \\
          &  &Circulation & \acrshort{rtpa} & SegNe &   & &\acrshort{fcv}&  \\ 
          \midrule 

        \multirow{2}{*}{\citet{Tolhuisen_2022}} & 
           \multirow{2}{*}{2022} & 
           
           \multirow{2}{*}{\acrshort{ais}} & 
           \multirow{1}{*}{\acrshort{evt}} & 
           \multirow{1}{*}{\acrshort{cae}+} & 
           \multirow{2}{*}{\acrshort{dwi}} & 
           \multirow{2}{*}{206} &
           \multirow{1}{*}{41 test} &
           \multirow{2}{*}{\acrshort{auc}: 0.88} \\
          &  & &\acrshort{tpa} & \acrshort{svm} &   & &cases&  \\ 
          \midrule

        \multirow{2}{*}{\citet{Moulton_2022}} & 
           \multirow{2}{*}{2022} & 
   
           \multirow{2}{*}{\acrshort{lvo}} & 
           \multirow{1}{*}{\acrshort{evt}} & 
           \multirow{1}{*}{\acrshort{cnn} +} & 
           \multirow{2}{*}{\acrshort{dwi}} & 
           \multirow{2}{*}{322} &
           \multirow{2}{*}{\acrshort{loocv} } &
           \multirow{2}{*}{\acrshort{auc}: 0.83} \\
          &  & & \acrshort{tpa} & Attention &   & &    &  \\ 
          \midrule

        \multirow{2}{*}{\citet{Xia_2023}} & 
           \multirow{2}{*}{2023} & 

           \multirow{2}{*}{\acrshort{sich}} & 
           \multirow{2}{*}{\acrshort{nr}} & 
           \multirow{2}{*}{\acrshort{cnn}} & 
           \multirow{2}{*}{\acrshort{ncct}} & 
           \multirow{2}{*}{377} &
           \multirow{2}{*}{91 ext. val } &
           \multirow{2}{*}{\acrshort{acc}: 0.706} \\
          &  & & &  &   & &   &  \\ 
          \midrule 
          
        \multirow{2}{*}{\citet{marcus2023stroke}} & 
           \multirow{2}{*}{2023} & 

           \multirow{2}{*}{\acrshort{ais}} & 
           \multirow{2}{*}{\acrshort{nr}} & 
           \multirow{1}{*}{{ResNet}} & 
           \multirow{2}{*}{\acrshort{ct}} & 
           \multirow{2}{*}{3573} &
           \multirow{1}{*}{80:20 } &
           \multirow{2}{*}{\acrshort{acc}: 0.788} \\
          &  & & & DDPM  &   & &  5-fold \acrshort{fcv} &  \\ 

          \midrule 
          
        \multirow{2}{*}{\citet{Yang_2024}} & 
           \multirow{2}{*}{2024} & 

           \multirow{2}{*}{\acrshort{ais}} & 
           \multirow{1}{*}{\acrshort{evt}} & 
           \multirow{2}{*}{{ResNet50}} & 
           \multirow{1}{*}{\acrshort{dwi}} & 
           \multirow{2}{*}{3338} &
           \multirow{2}{*}{60:20:20 } &
           \multirow{2}{*}{\acrshort{auc}: 0.788} \\
          &  & &\acrshort{tpa}&  & \acrshort{adc}  & & &  \\ 
         \midrule 
          
        \multirow{2}{*}{\citet{yang2024ischemic}} & 
           \multirow{2}{*}{2024} & 

           \multirow{2}{*}{\acrshort{ais}} & 
           \multirow{2}{*}{\acrshort{nr}} & 
           \multirow{1}{*}{\acrshort{cnn},\acrshort{lstm}} & 
           \multirow{2}{*}{\acrshort{dscpwi}} & 
           \multirow{2}{*}{88} &
           \multirow{2}{*}{10-fold \acrshort{fcv}} &
           \multirow{1}{*}{Mean Score: } \\
          &  & & & \acrshort{rnn}  &  & & &0.971  \\

        \bottomrule
    \end{tabular}

}
\caption{Overview of the studies that use only imaging information to predict functional outcome (\acrshort{mrs} scores) as stroke treatment outcome. The abbreviations are as follows; {\footnotesize \acrfull{ais}, \acrfull{lvo},  \acrfull{evt}, \acrfull{sich}, \acrfull{tpa}, \acrfull{rtpa}, \acrfull{cnn}, \acrfull{cae}, \acrfull{svm}, Denoising Diffusion Probabilistic Model (DDPM), \acrfull{cta}, \acrfull{dwi}, \acrfull{mpmri}, \acrfull{mri}, \acrfull{ncct}, \acrfull{ct},\acrfull{dscpwi}, \acrfull{loocv}, \acrfull{fcv}, \acrfull{auc}, \acrfull{acc}, \acrfull{nr}}.}
\label{table:outcome_mrs_image}
\end{table}

\begin{figure}[!t] %[ht]
    \centering
    \includegraphics[width=0.9\linewidth]{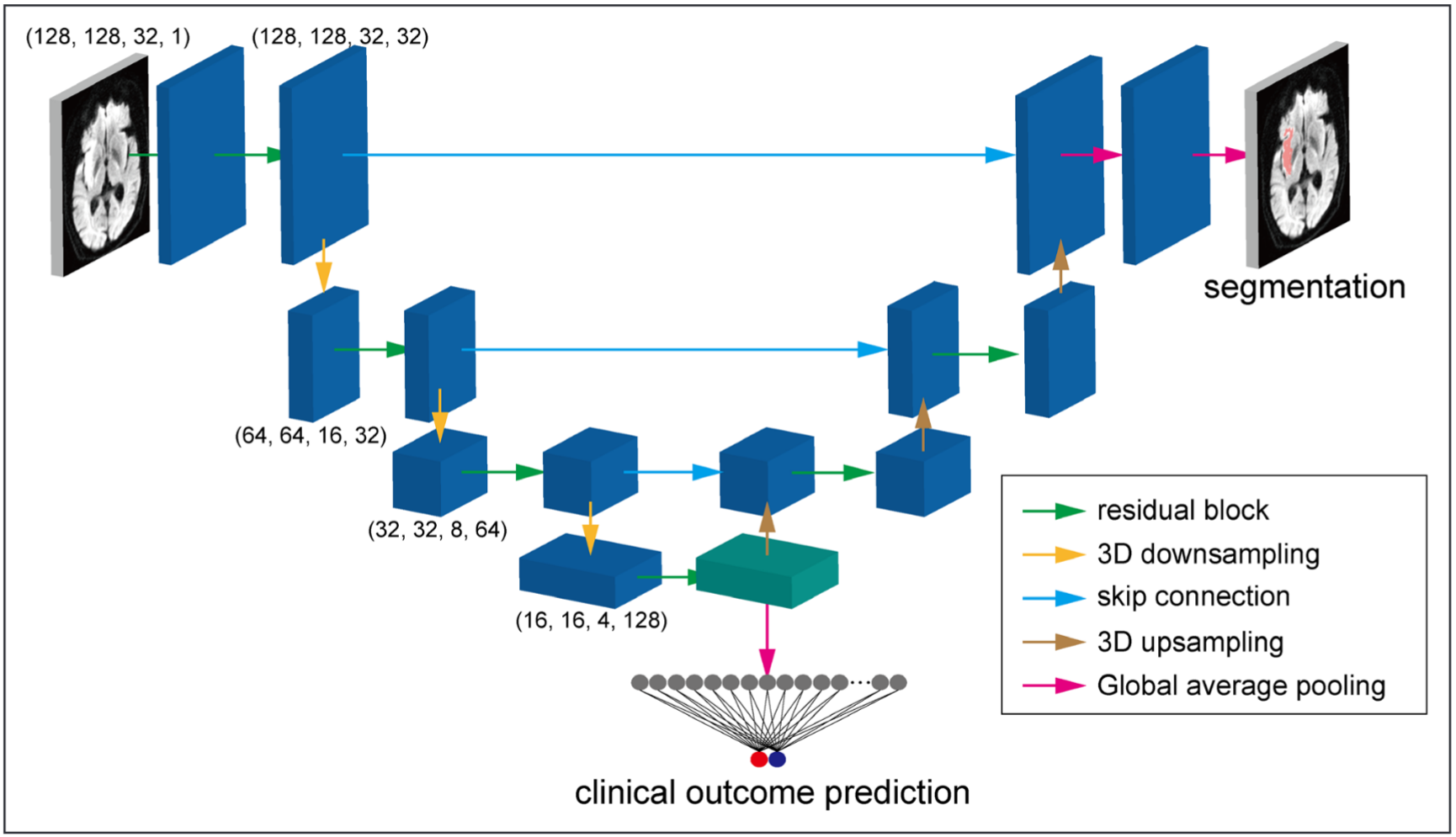}
    \caption{A multi-task model introduced by \cite{nishi2020deep} for lesion segmentation and clinical outcome prediction. {In the bottleneck of the U-Net model, binarised \gls{mrs} scores are predicted and the output of the decoder is the lesion core (indicated in red).} Figure from \cite{nishi2020deep}.}
    \label{figc2:nishi_network}
\end{figure}

\citet{Osama2020PredictingNetwork} proposed a parallel multi-parametric feature embedded Siamese network (PMFE-SN) and achieved high accuracy even with a small number of samples (an accuracy of 0.67 with just two minority class samples for training data of the ISLES 2017 dataset \cite{Winzeck2018ISLESMRI}) and effectively handled class imbalance in their \gls{mri} data.  \citet{Lai_2022} utilised a VGG-16 network trained on an \gls{mri} dataset of 44 patients, achieving an accuracy of 0.932 and 0.927 in the prediction of \gls{mrs} and \gls{nihss} respectively. {Their dataset was collected from two hospitals in China and also available by request from the authors}.   \citet{Moulton_2022} demonstrated that a \gls{cnn} with an attention mechanism, trained on 322 \gls{dwi} scans of one day post-stroke patients, gathered from ASTER \cite{lapergue2017effect} and INSULINFARCT \cite{rosso2012intensive} trials and the Pitié-Salpêtrière registry \cite{rosso2019impact},  achieved  better performance (\gls{auc} of 0.83) compared to {\gls{lr}} models using lesion volume (\gls{auc} of 0.78) or \gls{aspects} (\gls{auc} of 0.77) in predicting long-term functional outcomes at three months post-stroke.

Other researchers have explored \gls{dl} to extract more nuanced information from imaging data. \citet{Tolhuisen_2022} investigated whether information beyond infarct volume from follow-up \gls{dwi} could improve the prediction of the outcome in patients with acute ischaemic stroke.  They  found that classifiers using features extracted by a \gls{cae} or radiomics from follow-up \gls{dwi} scans (extracted from HERMES \cite{goyal2016endovascular}, ISLES 2015 \cite{maier2017isles}, and \gls{mrclean}-NO IV \cite{lecouffe2021randomized} datasets), achieved higher \gls{auc} values (0.88 and 0.81, respectively) compared to a model based solely on infarct volume (\gls{auc} of 0.79). \citet{marcus2023stroke} introduced a new approach that utilised a Denoising Diffusion Probabilistic Model (DDPM) to capture the evolution of stroke for predicting \gls{mrs} and \gls{nihss}  at discharge from \gls{ct} images in a dataset\footnote{\url{https://www.hra.nhs.uk/planning-and-improving-research/application-summaries/research-summaries/prediction-of-stroke-outcome-using-brain-imaging-machine-learning/}} of 3573 \gls{ais} patients, based solely on initial patient scans. They also incorporated both longitudinal data and time since stroke onset and achieved {\gls{auc} of 0.788 in \gls{mrs} and} \gls{auc} of 0.669 in next-day stroke severity (\gls{nihss}). 

\subsubsection{Multimodal Data Analysis }
\label{sec:outcome_multimodal}

Studies in the previous section have demonstrated the potential of using imaging data alone to predict stroke treatment outcomes. However, incorporating additional data of other modalities, such as neuroimaging data, imaging biomarkers and clinical information, has been shown to improve the accuracy of outcome prediction \cite{samak2020prediction,Ding_2021,Herzog_2023}. These multimodal  (fusion) models capture complementary information from various data sources, that can potentially leading to better performance \cite{Bacchi_2020,Liu_2023,zihni2020multimodal}.

The {outcome prediction} studies using multimodal data are listed in Table \ref{table:outcome_mrs_image_clinic}. {These studies employ various \gls{dl} architectures,} ranging from \gls{cnn}, \gls{lstm} to Transformers. {The models utilise diverse data sources, including} imaging information (\eg  \gls{ncct}, \gls{cta}, \gls{ctp})  and clinical variables (\eg demographics of the patients such as age and sex, and medical records detailing specifying like diabetes, hypertension and medications, etc.).

\begin{table}[ht]
\renewcommand{\arraystretch}{0.87}
\centering
\resizebox{\textwidth}{!}{

    \begin{tabular}{lclllcccll}
        \toprule
        % \specialrule{.2em}{.1em}{.1em}

        \multirow{2}{*}{\textbf{Study}} & 
           \multirow{2}{*}{\textbf{Year}}  & 
           \multirow{1}{*}{\textbf{Patient}} &
           \multirow{1}{*}{\textbf{Treatment}} & 
           \multirow{2}{*}{\textbf{Method}} & 
           \multirow{1}{*}{\textbf{Image}} & 
           \multirow{1}{*}{\textbf{Clinical}} & 
           \multirow{1}{*}{\textbf{Number of}} & 
           \multirow{1}{*}{\textbf{Data}} &
           \multirow{1}{*}{\textbf{Best}} \\
          &   & \textbf{Subgroup}  & \textbf{Type}  & & \textbf{Modality }  & \textbf{Features}& \textbf{Patients} & \textbf{Split} & \textbf{Result}  \\ 
        \midrule

       \multirow{2}{*}{\citet{choi2016ensemble}} & 
           \multirow{2}{*}{2016} & 

           \multirow{2}{*}{\acrshort{ais}} & 
           \multirow{2}{*}{\acrshort{nr}} & 
           \multirow{1}{*}{U-Net } & 
           \multirow{1}{*}{\acrshort{mri}} & 
           \multirow{2}{*}{3} &
           \multirow{2}{*}{49} &
           \multirow{1}{*}{19 test} &
           \multirow{2}{*}{\acrshort{auc}: 0.74} \\
          & & & &+ FC  & Sequences  && &cases &  \\ 
          \midrule 

        \multirow{2}{*}{\citet{Bacchi_2020}} & 
           \multirow{2}{*}{2020} & 

           \multirow{2}{*}{\acrshort{ais}} & 
           \multirow{2}{*}{\acrshort{tpa}} & 
           \multirow{1}{*}{\acrshort{cnn}+} & 
           \multirow{2}{*}{\acrshort{ncct}} & 
           \multirow{2}{*}{\acrshort{ns}} & 
           \multirow{2}{*}{204} &
           \multirow{2}{*}{85:15 } &
           \multirow{2}{*}{\acrshort{auc}: 0.75} \\
          & & & & \acrshort{ann} &   & & &    &  \\ 
          \midrule 
        
        \multirow{2}{*}{\citet{samak2020prediction}*} & 
           \multirow{2}{*}{2020} & 

           \multirow{2}{*}{\acrshort{is}} & 
           \multirow{2}{*}{\acrshort{evt}} & 
           \multirow{1}{*}{\acrshort{cnn}+} & 
           \multirow{2}{*}{\acrshort{ncct}} & 
           \multirow{2}{*}{27} & 
           \multirow{2}{*}{500} &
           \multirow{2}{*}{80:20} &
           \multirow{2}{*}{\acrshort{auc}: 0.75} \\
          & & & & Attention &   & & &    &  \\ 
          \midrule 
         \multirow{2}{*}{\citet{Heo_2020}} & 
           \multirow{2}{*}{2020} & 
  
           \multirow{2}{*}{\acrshort{ais}} & 
           \multirow{2}{*}{\acrshort{nr}} & 
           \multirow{1}{*}{\acrshort{cnn}+} &            
           \multirow{1}{*}{Textual} & 
           \multirow{1}{*}{\acrshort{mri}} & 
           \multirow{2}{*}{1840} &
           \multirow{2}{*}{70:30 } &
           \multirow{2}{*}{\acrshort{auc}: 0.75} \\
          &  & & & \acrshort{nlp} & Features & Reports & &    &  \\ 
          \midrule 

        \multirow{2}{*}{\citet{zihni2020multimodal}*} & 
           \multirow{2}{*}{2020} & 

           \multirow{1}{*}{Acute} & 
           \multirow{2}{*}{\acrshort{nr}} & 
           \multirow{1}{*}{\acrshort{cnn}+} & 
           \multirow{2}{*}{\acrshort{mra}} & 
           \multirow{2}{*}{7} & 
           \multirow{2}{*}{316} &
           \multirow{1}{*}{63 test} &
           \multirow{2}{*}{\acrshort{auc}: 0.76} \\
          & &cerebrovascular & & \acrshort{mlp} &   & & &  cases  &  \\ 
          \midrule 

        \multirow{2}{*}{\citet{Ding_2021}} & 
           \multirow{2}{*}{2021} & 
           
           \multirow{1}{*}{Acute} & 
           \multirow{1}{*}{Reperfusion} & 
           \multirow{2}{*}{\acrshort{cnn}} & 
           \multirow{1}{*}{\acrshort{dwi}} & 
           \multirow{2}{*}{\acrshort{ns}} & 
           \multirow{2}{*}{1438} &
           \multirow{2}{*}{75:25} &
           \multirow{2}{*}{\acrshort{auc}: 0.975} \\
          &  & Brainstem &Therapy &  & \acrshort{adc} & & &    &  \\ 
          \midrule 
          
        \multirow{2}{*}{\citet{Samak_2022}*} & 
           \multirow{2}{*}{2022} & 
 
           \multirow{2}{*}{\acrshort{ais}} & 
           \multirow{2}{*}{\acrshort{evt}} & 
           \multirow{2}{*}{\acrshort{cnn}} & 
           \multirow{2}{*}{\acrshort{ncct}} & 
           \multirow{2}{*}{27} & 
           \multirow{2}{*}{500} &
           \multirow{2}{*}{75:25 } &
           \multirow{2}{*}{\acrshort{auc}: 0.82} \\
          &  & & &  &   & & &    &  \\ 
          \midrule

        \multirow{2}{*}{\citet{Ramos_2022}*} & 
           \multirow{2}{*}{2022} & 
     
           \multirow{2}{*}{\acrshort{ais}} & 
           \multirow{2}{*}{\acrshort{evt}} & 
           \multirow{1}{*}{\acrshort{cnn}} & 
           \multirow{2}{*}{\acrshort{cta}} & 
           \multirow{2}{*}{50} & 
           \multirow{2}{*}{3279} &
           \multirow{1}{*}{5-fold } &
           \multirow{2}{*}{\acrshort{auc}: 0.81} \\
          &  & & & (ResNet10)  &   & & & \acrshort{fcv} &  \\ 
          \midrule

        \multirow{2}{*}{\citet{Hatami_2022}} & 
           \multirow{2}{*}{2022} & 
 
           \multirow{1}{*}{Proximal} & 
           \multirow{2}{*}{\acrshort{evt}} & 
           \multirow{1}{*}{\acrshort{cnn}+} & 
           \multirow{1}{*}{\acrshort{mri}} & 
           \multirow{2}{*}{4} & 
           \multirow{2}{*}{119} &
           \multirow{1}{*}{5-fold } &
           \multirow{2}{*}{\acrshort{auc}: 0.77} \\
          &  &occlusion & & \acrshort{lstm} & Sequences  & & &\acrshort{fcv}    &  \\ 
          \midrule 
        \multirow{2}{*}{\citet{Liu_2023}} & 
           \multirow{2}{*}{2023} & 

           \multirow{2}{*}{\acrshort{ais}} & 
           \multirow{1}{*}{\acrshort{evt}} & 
           \multirow{2}{*}{\acrshort{cnn}} & 
           \multirow{2}{*}{\acrshort{dwi}} & 
           \multirow{2}{*}{13} &
           \multirow{2}{*}{640} &
           \multirow{1}{*}{70:15:15 } &
           \multirow{2}{*}{\acrshort{auc}: 0.92} \\
          &  & &\acrshort{tpa}&  &   & &&280 ext. val.&  \\ 
          \midrule 
          
        \multirow{2}{*}{\citet{Herzog_2023}*} & 
           \multirow{2}{*}{2023} & 

           \multirow{2}{*}{\acrshort{mca}} & 
           \multirow{2}{*}{\acrshort{evt}} & 
           \multirow{1}{*}{\acrshort{cnn}+} & 
           \multirow{1}{*}{\acrshort{dwi},} & 
           \multirow{2}{*}{32} & 
           \multirow{2}{*}{222} &
           \multirow{1}{*}{50 test} &
           \multirow{2}{*}{\acrshort{auc}: 0.766} \\
          &  & & & \acrshort{ann} & \acrshort{pwi}& & &cases &  \\ 
          \midrule 

        \multirow{2}{*}{\citet{Chen_2023}*} & 
           \multirow{2}{*}{2023} & 
 
           \multirow{2}{*}{\acrshort{mca}} & 
           \multirow{2}{*}{\acrshort{tpa}} & 
           \multirow{2}{*}{\acrshort{cnn}+} & 
           \multirow{2}{*}{\acrshort{ctp}} & 
           \multirow{2}{*}{10} & 
           \multirow{2}{*}{230} &
           \multirow{1}{*}{80:20} &
           \multirow{2}{*}{\acrshort{auc}: 0.865} \\
          &  & & &  &   & & & patients   &  \\ 
          \midrule 

        \multirow{2}{*}{\citet{Hung_2023}} & 
           \multirow{2}{*}{2023} & 

           \multirow{2}{*}{\acrshort{ich}} & 
           \multirow{2}{*}{\acrshort{nr}} & 
           \multirow{1}{*}{\acrshort{nlp}} & 
           \multirow{2}{*}{EHR} & 
           \multirow{2}{*}{\acrshort{ns}} & 
           \multirow{2}{*}{1363} &
           \multirow{2}{*}{75:25 } &
           \multirow{2}{*}{\acrshort{auc}: 0.914} \\
          &  & & & \acrshort{dl}, \acrshort{ml} &   & & &    &  \\ 
          \midrule 
                
        \multirow{2}{*}{\citet{Shen_2023}} & 
           \multirow{2}{*}{2023} & 
 
           \multirow{2}{*}{\acrshort{ais}} & 
           \multirow{2}{*}{\acrshort{evt}} & 
           \multirow{2}{*}{\acrshort{cnn}} & 
           \multirow{2}{*}{\acrshort{cta}} & 
           \multirow{2}{*}{\acrshort{ns}} & 
           \multirow{2}{*}{44} &
           \multirow{2}{*}{\acrshort{ns}} &
           \multirow{2}{*}{\acrshort{auc}: 0.874} \\
          &  & & &  &   & & &    &  \\ 
          \midrule 

        \multirow{2}{*}{\citet{Samak_2023}*} & 
           \multirow{2}{*}{2023} & 
    
           \multirow{2}{*}{\acrshort{ais}} & 
           \multirow{2}{*}{\acrshort{evt}} & 
           \multirow{1}{*}{Transformer,} & 
           \multirow{2}{*}{\acrshort{ncct}} & 
           \multirow{2}{*}{27} & 
           \multirow{2}{*}{500} &
           \multirow{2}{*}{75:25 } &
           \multirow{2}{*}{\acrshort{auc}: 0.85} \\
          &  & & & \acrshort{cnn} &   & & &    &  \\ 

        \midrule 

        \multirow{2}{*}{\citet{zeng2023improved}} & 
           \multirow{2}{*}{2023} & 
    
           \multirow{2}{*}{\acrshort{is}} & 
           \multirow{2}{*}{\acrshort{evt}} & 
           \multirow{1}{*}{\acrshort{cnn}} & 
           \multirow{1}{*}{\acrshort{ncct},\acrshort{cta}} & 
           \multirow{2}{*}{6} & 
           \multirow{2}{*}{460} &
           \multirow{1}{*}{4-fold} &
           \multirow{2}{*}{\acrshort{auc}: 0.807} \\
          &  & & & L2GAN & \acrshort{ctp}  & & & \acrshort{fcv} &  \\ 

        \midrule 

        \multirow{2}{*}{\citet{Oliveira2023}*} & 
           \multirow{2}{*}{2023} & 
    
           \multirow{2}{*}{\acrshort{is}} & 
           \multirow{1}{*}{\acrshort{evt}} & 
           \multirow{1}{*}{\acrshort{cnn}} & 
           \multirow{2}{*}{\acrshort{ncct}} & 
           \multirow{2}{*}{8} & 
           \multirow{2}{*}{743} &
           \multirow{1}{*}{10-folds } &
           \multirow{2}{*}{\acrshort{auc}: 0.806} \\
          & & & \acrshort{tpa}  & Siamese &   & & & \acrshort{fcv} &  \\ 

        \midrule 

        \multirow{2}{*}{\citet{Borsos2024}} & 
           \multirow{2}{*}{2023} & 
    
           \multirow{2}{*}{\acrshort{ais}} & 
           \multirow{2}{*}{\acrshort{nr}} & 
           \multirow{1}{*}{ResNet,} & 
           \multirow{2}{*}{\acrshort{ctp}} & 
           \multirow{2}{*}{14} & 
           \multirow{2}{*}{98} &
           \multirow{2}{*}{60:20:20 } &
           \multirow{2}{*}{\acrshort{auc}: 0.75} \\
          &  & & & DAFT \cite{polsterl2021combining} &   & & &    &  \\ 

        \midrule 
        \multirow{2}{*}{\citet{Jo_2023Comnining}} & 
           \multirow{2}{*}{2023} & 

           \multirow{2}{*}{\acrshort{ais}} & 
           \multirow{2}{*}{\acrshort{nr}} & 
           \multirow{2}{*}{DenseNet} & 
           \multirow{1}{*}{\acrshort{dwi}} & 
           \multirow{2}{*}{3} & 
           \multirow{2}{*}{4147} &
           \multirow{1}{*}{80:20} &
           \multirow{2}{*}{\acrshort{auc}: 0.786} \\
          &  & & &  & \acrshort{adc}  & & &5-fold \acrshort{fcv}&  \\ 
          \midrule 

        \multirow{2}{*}{\citet{jung2024multimodal}} & 
           \multirow{2}{*}{2024} & 

           \multirow{2}{*}{\acrshort{ais}} & 
           \multirow{2}{*}{\acrshort{nr}} & 
           \multirow{1}{*}{ResNeXt} & 
           \multirow{1}{*}{\acrshort{dwi}} & 
           \multirow{2}{*}{22} & 
           \multirow{2}{*}{2606} &
           \multirow{1}{*}{5-fold} &
           \multirow{2}{*}{\acrshort{auc}: 0.83} \\
          &  & & & + CBAM & \acrshort{flair}  & & &    \acrshort{fcv}&  \\ 
        \midrule

        \multirow{2}{*}{\citet{Diprose_2024}*} & 
           \multirow{2}{*}{2024} & 

           \multirow{2}{*}{\acrshort{ais}} & 
           \multirow{2}{*}{\acrshort{evt}} & 
           \multirow{1}{*}{DenseNet121} & 
           \multirow{2}{*}{\acrshort{ncct}} & 
           \multirow{2}{*}{26} & 
           \multirow{2}{*}{975} &
           \multirow{2}{*}{80:20} &
           \multirow{2}{*}{\acrshort{auc}: 0.811} \\
          &  & & & \acrshort{lr} &   & & & &  \\ 
          \midrule 

        \multirow{2}{*}{\citet{Liu_2024}} & 
           \multirow{2}{*}{2024} & 

           \multirow{2}{*}{\acrshort{ais}} & 
           \multirow{2}{*}{\acrshort{nr}} & 
           \multirow{2}{*}{\acrshort{dl}} & 
           \multirow{2}{*}{\acrshort{dwi}} & 
           \multirow{2}{*}{\acrshort{ns}} & 
           \multirow{2}{*}{80} &
           \multirow{2}{*}{\acrshort{ns}} &
           \multirow{2}{*}{\acrshort{acc}: 0.36} \\
          &  & & &  &   & & &    &  \\ 
          \midrule 

        \multirow{2}{*}{\citet{Mart_n_Vicario_2024}*} & 
           \multirow{2}{*}{2024} & 

           \multirow{2}{*}{\acrshort{ais}} & 
           \multirow{2}{*}{\acrshort{evt}} & 
           \multirow{1}{*}{\acrshort{gcn}} & 
           \multirow{1}{*}{\acrshort{cta}} & 
           \multirow{2}{*}{81} & 
           \multirow{2}{*}{220} &
           \multirow{1}{*}{5-fold} &
           \multirow{2}{*}{\acrshort{auc}: 0.87} \\
          &  & & &\acrshort{fcn}& \acrshort{ctp}  & & &    \acrshort{fcv}&  \\ 
          \midrule 
          \multirow{2}{*}{\citet{liu2024prediction}} & 
           \multirow{2}{*}{2024} & 

           \multirow{2}{*}{\acrshort{ais}} & 
           \multirow{2}{*}{\acrshort{evt}} & 
           \multirow{2}{*}{ResNet} & 
           \multirow{2}{*}{\acrshort{ncct}} & 
           \multirow{2}{*}{7} & 
           \multirow{2}{*}{1335} &
           \multirow{1}{*}{6-fold} &
           \multirow{2}{*}{\acrshort{auc}: 0.91} \\
          &  & & & &   & & &    \acrshort{fcv}&  \\ 
\midrule 
          \multirow{2}{*}{\citet{amador2024cross}*} & 
           \multirow{2}{*}{2024} & 

           \multirow{2}{*}{\acrshort{ais}} & 
           \multirow{2}{*}{\acrshort{evt}} & 
           \multirow{1}{*}{CNN} & 
           \multirow{2}{*}{\acrshort{ctp}} & 
           \multirow{2}{*}{9} & 
           \multirow{2}{*}{70} &
           \multirow{1}{*}{10-fold} &
           \multirow{2}{*}{\acrshort{auc}: 0.77} \\
          &  & & &Cross-Attention&   & & &    \acrshort{fcv}&  \\

        \bottomrule
    \end{tabular}
}
\caption{Overview of the studies that perform functional outcome as stroke treatment outcome and use imaging and clinical information. The abbreviations are as follows; {\tiny \acrfull{is}, \acrfull{ais}, \acrfull{mca}, \acrfull{ich},  \acrfull{evt}, \acrfull{tpa}, \acrfull{dl}, \acrfull{ml}, \acrfull{cnn}, \acrfull{ann}, \acrfull{mlp}, \acrfull{nlp}, \acrfull{lstm}, \acrfull{mri}, \acrfull{ctp}, \acrfull{dwi}, \acrfull{pwi}, \acrfull{ncct}, \acrfull{cta}, \acrfull{adc}, Electronic Health Records (EHR), \acrfull{fcv}, \acrfull{auc}, \acrfull{nr}, \acrfull{ns}. {* indicates that the code used in the study is available.}}}
\label{table:outcome_mrs_image_clinic}
\end{table}

\citet{choi2016ensemble} proposed an ensemble of deep neural networks that achieved \gls{auc} of 0.74 on ISLES 2016 \cite{Winzeck2018ISLESMRI} dataset in clinical outcomes  after ischaemic stroke treatment. \citet{samak2020prediction} used a multimodal \gls{cnn} approach, incorporating an attention mechanism to capture inter-dependencies among features, to predict the success of endovascular treatment for ischaemic stroke patients from \gls{ncct} acquired at the hospital admission. Their model achieved \gls{auc} of 0.75 in dichotomised \gls{mrs} scores and accuracy of 0.35 in individual \gls{mrs} scores prediction on \gls{mrclean} dataset. \citet{Liu_2023} developed a fusion model that combined \gls{dwi} and clinical features extracted using a ResNet and \gls{svm} model respectively, to predict 90-day functional outcomes. Their model achieved 0.92 \gls{auc} on a dataset of 640 patients that included four multi-centre trials {(iCAS, DEFUSE-2, DEFUSE-3, CRISP)} and two single center registries {(University of California, Los Angeles stroke registry and Lausanne University Hospital stroke registry)}, demonstrating better performance than models based imaging features (\gls{auc} of 0.88) or solely on clinical (\gls{auc} of 0.88). \citet{Chen_2023} used a \gls{cnn} {which included five convolutional layers along with an efficient channel attention (ECA) layer \cite{wang2020eca}} to encode features of \gls{ctp} maps and combined them with demographic data features to improve post-thrombolysis functional outcome prediction. Their model was trained on 230 patients and validated on 129 patients, and achieved an \gls{auc} of 0.865 in a multimodal configuration, surpassing the performance of models that relied only on imaging data (\gls{auc} of 0.792) and clinical data (\gls{auc} of 0.670). {No information is provided on the availability of their dataset.}

In  \cite{Bacchi2019DeepStudy}, Bacchi \etal used a dataset consisting of 204 patients with \gls{ncct} image volumes and relevant clinical information, such as age, gender, blood pressure, and \gls{nihss} scores,  to predict the binary outcomes of thrombolysis treatment with both \gls{mrs} at 90 days and \gls{nihss} at 24 hours. {They found that their most successful model was a combination of {a custom} \gls{cnn} {comprising two 3D convolution layers followed by a maximum pooling layer and two \gls{fc} layers} and an \gls{ann} {based on 3 \gls{fc} layers and  ReLU activation function} that were trained on both  \gls{ncct} image volumes and  clinical metadata.
This model reached  an \gls{auc} of 0.75 in  \gls{mrs}  and  0.70 in \gls{nihss} score prediction. \citet{Bacchi2019DeepStudy} also did not make their data available.}

\citet{Ramos_2022} investigated the use of \gls{ml} and \gls{dl} to predict clinical outcomes in stroke patients undergoing \gls{evt}. They used both imaging and clinical data from the \gls{mrclean} registry and explored two approaches: one using radiomics features extracted from specific brain regions and the other employing 3D \gls{dl} models analysing whole brain images combined with clinical information by using \gls{ml} (\gls{rf}, \gls{svm}, \gls{xgb}, \gls{ann}) and \gls{dl} (ResNet10) respectively. Interestingly, contrary to other studies, incorporating imaging data did not significantly improve the performance of \gls{mrs} prediction. {Using only clinical data, \gls{auc} was 0.81 for \gls{ml} and 0.77 for \gls{dl}, while when using both imaging and clinical data, the \gls{auc} remained at 0.80 for \gls{ml} and 0.77 for \gls{dl}.} However, for \gls{tici} score prediction, the inclusion of imaging data increased the \gls{auc} from 0.53 to 0.57 for \gls{ml} and from 0.53 to 0.61 for \gls{dl}.

Early prediction (at one week) of stroke evolution has been found to be critical for predicting the functional outcome after stroke treatment. \citet{ernst2017association} highlighted this by demonstrating a correlation between the volume of lesions in 1-week follow-up NCCT scans with functional outcome. Building on this, \citet{Samak_2022}  developed a model to predict \gls{mrs} scores by incorporating follow-up scan data.  Their approach, named Feature Matching Auto-encoder (FeMA) (see Fig \ref{figc2:fema_network}), leveraged a self-supervised voxel-wise method based on {a custom} \gls{cnn} {model that consisted of three encoder and a decoder modules}  to predict follow-up scans in stroke patients from their baseline scan {alone}. In training their model,  a two-step  strategy was used where in the first stage a 1-week follow-up scan was reconstructed from a baseline scan to encode information about tissue changes over a week after treatment, and then in the second stage, the learnt features from the first stage and a baseline scan were combined to estimate \gls{mrs} scores. The model was tested using the \gls{mrclean} trial dataset and obtained an \gls{auc} of 0.82.

\begin{figure}[!t]
    \centering
    \includegraphics[width=\linewidth]{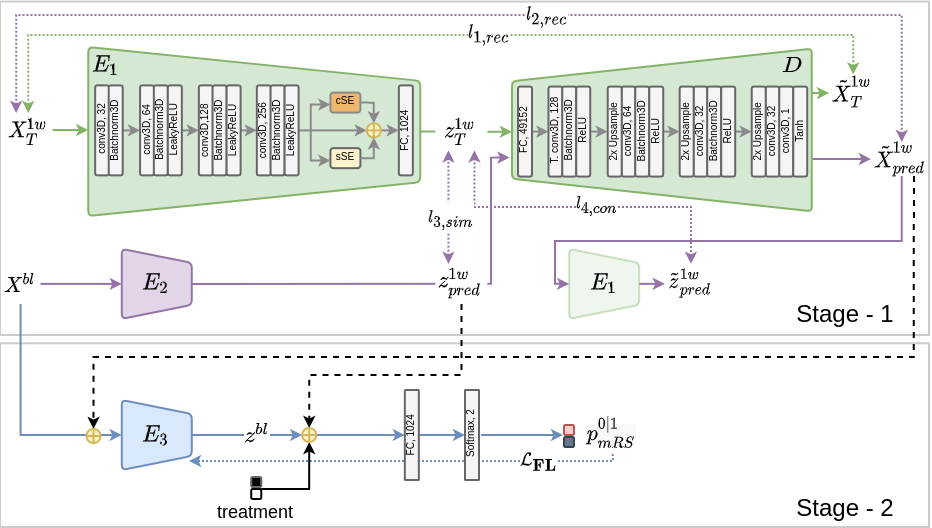}
    \caption{Overview of the FeMA network \cite{Samak_2022}. The model predicts the follow-up \acrshort{ncct} scan (1-week follow-up $X^{1w}_{T}$) from the baseline scan $X^{bl}$. Then, it combines the baseline scan with the predicted follow-up scan $X^{1w}_{pred}$ and 1-week follow-up scan features $z^{1w}_{pred}$ to predict the functional outcome of the stroke treatment. Figure from \cite{Samak_2022}.}
    \label{figc2:fema_network}
\end{figure}

Interpretable deep learning models have also been developed to improve the estimation of functional outcomes. \citet{Herzog_2023} used an interpretable \gls{dl} model based on the study of \cite{kook2022deep} and demonstrated that utilising both imaging data (\gls{dwi}) and clinical variables (\eg age, systolic blood pressure, diabetes, hypertension, smoking). {They collected a dataset of 222 patients from Inselspital Bern, Switzerland which is available by requests from the authors. Their model} achieved an accuracy of 0.72, outperforming models based only on imaging data (accuracy of 0.614) or clinical data (accuracy of 0.60). The inclusion of imaging data allowed the model to significantly outperform experienced stroke neurologists, who had an accuracy of {0.64 with imaging and clinical data, an accuracy of 0.55 with imaging data only and an accuracy of 0.60 with clinical data alone.}  

A recent study by \citet{Mart_n_Vicario_2024} introduced an uncertainty-aware graph deep learning model to predict long-term functional outcomes and mortality rates in \gls{ais} patients following \gls{evt}. Their dataset was collected from the University Hospital of Erlangen, comprising 220 \gls{ais} patients, {and is available by request from the authors}. They employed \glspl{gcn} and \glspl{fcn}  to predict the outcome at admission, post-\gls{evt}, and 24 hours after stroke. Their results demonstrated comparable performance across the algorithms {including \gls{lr}, \gls{rf}, \gls{xgb} and \gls{fcn}}, with a maximum \gls{auc} of 0.87 for the prediction of \gls{mrs} using \glspl{gcn}. The performance of the predictions increased when data from later time points were incorporated, rising from an \gls{auc} of 0.76 at admission to 0.84 post‑\gls{evt} and to 0.87 at 24 hours post-stroke. 

In addition to using imaging and clinical data, \gls{nlp} has been used in several medical domains to analyse clinical narratives {\cite{sloan2024automated}  and applied to} treatment outcome prediction. By generating text-based markers from free-text \gls{mri} reports, \gls{nlp}-based \gls{ml} algorithms have demonstrated improved performance in predicting poor outcomes in \gls{ais} patients \cite{Heo_2020}. Similarly, \gls{nlp} techniques have been applied to enhance the prediction of functional outcomes after \gls{ich} \cite{Hung_2023}.

\section{Discussion}
\label{sec:discussion}

{So far in this article, we have reviewed the current landscape of stroke outcome prediction methods that have explored deep learning to make estimates at various time-points following a stroke. Next, we present our key observations with respect to the major issues and challenges, and discuss current limitations.} 

{\bf Use of Deep Learning --} A consistent feature across the reviewed studies is the dominance of using \gls{dl} models based on \glspl{cnn}, {for example \cite{Nielsen_2018,robben2020prediction,Osama2020PredictingNetwork,Samak_2022,amador2022predicting,marcus2023stroke,Palsson_2024,Diprose_2024}},  with U-Net variations commonly used for final infarct prediction \cite{choi2016ensemble,lucas2018learning,debs2021impact,amador2022predicting,Winder_2022,Palsson_2024}, due to their ability to handle segmentation tasks effectively, whereas ResNet-style architectures are preferred for predicting functional outcomes, {such as \cite{hilbert2019data,Fang_2022,Ramos_2022,Borsos2024,Yang_2024}}. Among imaging modalities, \gls{mri} is most used for both the prediction of the final infarct and functional outcome, as it provides detailed information about stroke tissue, {\eg \cite{Stier_2015,Nielsen_2018,pinto2021combining,Nazari_Farsani_2023}}. \gls{ctp} is popular for final infarct prediction due to its ability to assess cerebral blood flow, a key factor in tissue infarction  \cite{lucas2018learning,robben2020prediction,Winder_2022,Amador_2024}. \gls{ncct} is also frequently used for predicting functional outcomes, likely due to its greater availability and lower cost, {\eg \cite{Bacchi2019DeepStudy,samak2020prediction,Xia_2023,Samak_2023,Diprose_2024}}.

A common trend in the reviewed studies is the development of novel implementations rather than investigation and validation of existing methods \cite{Yeo2021}. Furthermore, many studies lack sufficient detail regarding model building procedures, data processing steps and hyperparameter tuning. In addition to this problem, only a small proportion of studies (15 out of 60) analysed in this research share their code publicly, such as \cite{samak2020prediction,pinto2021combining,amador2022hybrid,Chen_2023,Diprose_2024} as can be seen in Table \ref{tab:code_table}. Consequently, ensuring transparency and reproducibility remains a significant challenge. To address this, a commitment to open science practices is necessary. 

Finally, while \gls{dl} models have shown promising performance in predicting stroke outcomes, their lack of interpretability makes it difficult to assess their reliability and adoption in clinical settings \cite{Yeo2021}. Clinicians expect to better understand "black-box" DL models, so they can have confidence in and rely on their predictions \cite{Herzog_2023,Gutierrez_2024}. To address this critical gap, a few recent studies \cite{Moulton_2022,Herzog_2023,Gutierrez_2024,Mart_n_Vicario_2024} have explored interpretable \gls{dl} models in the context of \gls{sop}. 

{\bf Multimodal Data --} The majority of studies, particularly the studies on functional outcome prediction, leverage multimodal data, combining imaging and clinical information to achieve superior performance compared to models that rely solely on unimodal information \cite{Bacchi2019DeepStudy,Liu_2023,samak2020prediction,Samak_2023,Herzog_2023}. This approach acknowledges that the prediction of stroke outcomes depends not only on the extent of brain damage, but also on previous health conditions and other characteristics of the patient. Furthermore, incorporating follow-up (lesion outcome) information demonstrably improves the performance of models to predict \gls{mrs} scores, such as \cite{Samak_2022,Tolhuisen_2022}.

{\bf Availability of Data  \& Benchmarks --} Although \gls{dl} techniques such as \gls{cnn}s, and more recently transformer-based architectures,  hold significant promise for predicting stroke outcomes, several limitations need to be addressed. {{A significant challenge is the limited availability of large, all-encompassing, easily accessible and well-annotated datasets \cite{Karthik2020,Yeo2021} that can allow  \gls{dl} algorithms to learn complex patterns and ensure generalisability. The majority of the studies reviewed in this research use data sets consisting of fewer than 500 patients. This scarcity restricts the development of robust and generalisable models.}}

While significant efforts have been made to develop clinical trial datasets such as  \gls{mrclean} \cite{Berkhemer2015AStroke}, HERMES \cite{goyal2016endovascular}, ERASER \cite{fiehler2019eraser} and DIFUSE-2 \cite{lansberg2012mri}, are not structured as benchmarks for comparative evaluation of \gls{dl} models. This lack of standardised benchmarking framework (datawise and processes) has led to various niche datasets, inconsistent evaluation metrics and experimental setups, hence making it challenging to assess relative strengths and weaknesses between different algorithms fairly and objectively. Even when employing the same clinical trial dataset, inconsistencies arise due to the incomparable use of  data modalities and subsets for training and testing \eg \cite{hilbert2019data,Ramos_2022}. Clearly, collaborative efforts are needed to establish comprehensive standardised public datasets and evaluation protocols that allow researchers to perform thorough comparisons and accelerate progress in model development.

Although some clinical trial datasets offer data from multiple centres, many single-centre studies rely on in-house datasets, such as \cite{Hokkinen_2021,CHavva2022,Fang_2022,He_2022,yang2024ischemic}, which often lack heterogeneity. Stroke presentation, imaging equipment and neuroimaging parameters vary considerably across healthcare institutions, and this in-house approach fails to capture this essential diversity. Such lack of diversity in single-centre training data limits the generalisability of models to real-world scenarios with broader patient characteristics and clinical practices. It also leads to limited or no means of external validation which creates a significant obstacle to clinical translation \cite{Yeo2021}. Such lack of external validation, which has been explored by only a few  \cite{nishi2020deep,Wouters_2022,Xia_2023,Chen_2023}, is cause for concern on the reliability of reported results.

\section{Future Directions}
\label{sec:future_directions}

There is scope for several promising areas of research to significantly improve prediction accuracy and robustness for stroke outcome using deep learning.

\textbf{Adaptive Multimodal Data Fusion --} Combining information from various data sources, consistently improves the predictive performance of models \cite{Bacchi2019DeepStudy,Samak_2022,jung2024multimodal}. However, current methods often rely on naive fusion techniques, such as late fusion, which fail to explore the the complex relationships between different types of data~\cite{huang2020fusion,amador2024providing}. This suggests a critical need for the development of adaptive multimodal fusion methods that dynamically learn relationships and contextually integrate information, such as neuroimaging data and clinical records. 

\textbf{Leveraging Final Infarct Information --}
The information available in brain scans within post-treatment stroke lesions (final infarct) - specifically the size and location of the damaged area - is extremely valuable for predicting how well a patient will recover in the long term. While encoding final infarct information into models has shown promise for improving functional outcome prediction in some studies \cite{Samak_2022,Tolhuisen_2022}, this area still requires further investigation. Therefore, developing models that can effectively encode such information, along with longitudinal lesion changes, could enhance prediction robustness.

\textbf{Federated Learning --} Federated learning presents a promising solution to the challenge of data heterogeneity in developing robust models for \gls{sop}. This decentralised approach allows for collaborative model training across multiple institutions without compromising patient privacy by avoiding sharing source data \cite{guan2024federated}. By enabling the training of models on larger and more diverse datasets, federated learning facilitates the development of more generalisable models capable of handling the variability encountered in real-world clinical settings \cite{yan2023label}. Such collaborative approach holds significant potential for improving the performance and reliability of models in \gls{sop}s across diverse patient populations.

\textbf{Annotation-free Segmentation --} Currently, the reliance on manual lesion segmentation to generate ground truth labels in follow-up scans presents a major bottleneck due to its time-consuming and labour-intensive nature. The development of annotation-free models for final infarct segmentation offers a significant opportunity to improve efficiency and cost-effectiveness~\cite{Gutierrez_2024}. This would also allow researchers to access larger training datasets, leading to the development of more robust and accurate predictive models.

\section{Conclusions}
\label{sec:conclusion}
This paper has provided a review of deep learning approaches for stroke outcome prediction, encompassing both final infarct and functional outcomes, using imaging and multimodal data. We have presented the most recent techniques, datasets, evaluations, and code.   While significant progress has been made, particularly in utilising multimodal data and advanced model architectures, specifically U-Net variations for lesion outcome segmentation and ResNet architectures for functional outcome prediction, several critical challenges remain. The scarcity of large, standardised benchmark datasets and inherent data heterogeneity restrict the development of robust and generalisable models. Furthermore, the limited use of external validation, inconsistencies in reporting practices, and the lack of interpretable models raise concerns regarding the clinical translation and reliability of existing approaches.

Despite these challenges, several promising avenues for future research  are: the development of adaptive multimodal fusion methods that effectively leverage the complementary nature of diverse data sources,  the incorporation of longitudinal lesion features, the adoption of  federated learning techniques for improved data utilisation, and the exploration of annotation-free lesion labelling methods.

\section*{Declarations}

% Some journals require declarations to be submitted in a standardised format. Please check the Instructions for Authors of the journal to which you are submitting to see if you need to complete this section. If yes, your manuscript must contain the following sections under the heading `Declarations':

\begin{itemize}
\item \textbf{Funding } Not applicable.
\item \textbf{Conflict of interest } The authors declare no competing interests.
\item \textbf{Ethics approval and consent to participate } Not applicable.
% \item \textbf{Consent for publication } Not applicable.
% \item \textbf{Data availability }
% \item \textbf{Materials availability }
% \item \textbf{Code availability }
% \item \textbf{Author contribution }
\end{itemize}

\noindent
% If any of the sections are not relevant to your manuscript, please include the heading and write `Not applicable' for that section.

\begin{appendices}

\section{Studies with Code}\label{secA1}

\begin{table}[!ht]
    \centering
    \large
\caption{The list of the studies that have shared their code publicly.}
\label{tab:code_table}
\resizebox{\textwidth}{!}{
    \begin{tabular}{lll}
    \toprule
         \textbf{Study}& \textbf{Task} & \textbf{Code Link}\\
         \midrule
         \citet{lucas2018learning}& Final Infarct & \url{https://github.com/multimodallearning/stroke-prediction} \\
         \citet{samak2020prediction}& Functional Outcome & \url{https://github.com/zeynelsamak/Thrombectomy-Outcome}\\
         \citet{zihni2020multimodal} & Functional Outcome & \url{https://github.com/prediction2020/multimodal-classification}\\
         \citet{pinto2021combining}& Final Infarct  & \url{https://github.com/apint0/stroke_prediction}\\
         \citet{Samak_2022}& Functional Outcome  & \url{https://github.com/zeynelsamak/FeMA}\\
         \citet{Ramos_2022}& Functional Outcome & \url{https://github.com/L-Ramos/mrclean_combination}\\
         \citet{amador2022hybrid}& Final Infarct & \url{https://github.com/kimberly-amador/Spatiotemporal-CNN-Transformer}\\
         \citet{Lee_2022}& Final Infarct  & \url{https://github.com/sdlee087/deep_medical}\\
         \citet{Herzog_2023}& Functional Outcome  & \url{https://github.com/liherz/functional_outcome_prediction_dl_vs_neurologists}\\
         \citet{Chen_2023}& Functional Outcome & \url{https://github.com/Yutong441/deepCTP}\\
         \citet{Samak_2023}& Functional Outcome & \url{https://github.com/zeynelsamak/TranSOP}\\
         \citet{Oliveira2023} & Functional Outcome & \url{https://github.com/GravO8/mrs-dl} \\
         \citet{Diprose_2024} & Functional Outcome & \url{https://github.com/jdddog/deep-mt}\\
         \citet{Mart_n_Vicario_2024}& Functional Outcome& \url{https://gitlab.cs.fau.de/gu47bole/ais}\\
         \citet{amador2024cross}& Functional Outcome& \url{https://github.com/kimberly-amador/Multimodal-mRS90-Outcome-Prediction}\\   
    \bottomrule
    \end{tabular}
}
    
\end{table}

% An appendix contains supplementary information that is not an essential part of the text itself but which may be helpful in providing a more comprehensive understanding of the research problem or it is information that is too cumbersome to be included in the body of the paper.




\end{appendices}


\bibliography{sn-bibliography}% common bib file

\end{document}